\documentclass[mp,preprint,showpacs,floatfix]{revtex4}
\usepackage{amsmath}
\usepackage{amsfonts}
\usepackage{amssymb}
\usepackage{dcolumn}
\usepackage{bm}
\usepackage{graphicx}
\newcommand{\graphr}{
\setlength{\unitlength}{0.8em}
\begin{picture}(1.1,1)(0,-0.25)
\linethickness{0.1em}
\put(.5,.5){\circle*{.2}}
\linethickness{0.1em}
\put(0.5,0){\circle{1}}
\end{picture}
}

\newcommand{\graphgamma}{
\setlength{\unitlength}{0.8em}
\begin{picture}(1.5,2)(-0.25,-0.25)
\linethickness{0.1em}
\put(0,0){\circle*{.2}}
\put(1,0){\circle*{.2}}
\linethickness{0.05em}
\put(0,0){\line(1,0){1}}
\end{picture}
}

\newcommand{\graphrra}{
\setlength{\unitlength}{0.8em}
\begin{picture}(1.5,2)(-0.25,-0.25)
\linethickness{0.1em}
\put(0.5,0){\circle*{.2}}
\linethickness{0.1em}
\put(0.5,-0.5){\circle{1}}
\put(0.5,0.5){\circle{1}}
\end{picture}
}

\newcommand{\graphrrb}{
\setlength{\unitlength}{0.8em}
\begin{picture}(3,1.5)(-0.25,0)
\linethickness{0.1em}
\put(0.50,0.00){\circle*{.2}}
\put(0.50,0.50){\circle{1}}

\put(2,0.00){\circle*{.2}}
\put(2,0.50){\circle{1}}

\end{picture}
}

\newcommand{\graphgra}{
\setlength{\unitlength}{0.8em}
\begin{picture}(2.5,2)(-0.25,-0.25)
\linethickness{0.1em}
\put(1,0){\circle*{.2}}
\put(2,0){\circle*{.2}}
\put(0.5,0){\circle{1}}

\linethickness{0.05em}
\put(1,0){\line(1,0){1}}
\end{picture}
}

\newcommand{\graphgrb}{
\setlength{\unitlength}{0.8em}
\begin{picture}(3.5,2)(-0.25,-0.25)
\linethickness{0.1em}
\put(1,0){\circle*{.2}}
\put(2,0){\circle*{.2}}
\put(3,0){\circle*{.2}}
\put(0.5,0){\circle{1}}

\linethickness{0.05em}
\put(2,0){\line(1,0){1}}
\end{picture}
}

\newcommand{\graphgga}{
\setlength{\unitlength}{0.8em}
\begin{picture}(1.5,2)(-0.25,-0.25)
\linethickness{0.05em}
\put(0,0){\circle*{.2}}
\put(1,0){\circle*{.2}}
\linethickness{0.1em}
\put(0.5,0){\circle{1}}
\end{picture}
}

\newcommand{\graphggb}{
\setlength{\unitlength}{0.8em}
\begin{picture}(1.5,2)(-0.25,-0.25)
\linethickness{0.05em}
\put(0,-0.5){\circle*{.2}}
\put(0,-0.5){\line(1,0){1}}
\put(1,-.5){\circle*{.2}}
\put(1,-0.5){\line(0,1){1}}
\put(1,0.5){\circle*{.2}}
\end{picture}
}

\newcommand{\graphggc}{
\setlength{\unitlength}{0.8em}
\begin{picture}(1.5,2)(-0.25,-0.25)
\linethickness{0.05em}
\put(0,0.5){\line(1,0){1}}
\put(0,-0.5){\line(1,0){1}}
\linethickness{0.05em}
\put(0,0.5){\circle*{0.2}}
\put(0,-0.5){\circle*{0.2}}
\put(1,0.5){\circle*{0.2}}
\put(1,-0.5){\circle*{0.2}}
\end{picture}
}

\newcommand{\graphrrra}{
\setlength{\unitlength}{0.8em}
\begin{picture}(3,2)(-1.35,-0.25)
\linethickness{0.02em}
\put(0,0){\circle*{0.2}}
\qbezier(0.00,0.00)(0.281,1.875)(1.125,0.750)
\qbezier(1.125,0.750)(1.969,-0.375)(0.00,0.00)
\qbezier(0.00,0.00)(-1.406,-1.125)(0.00,-1.125)
\qbezier(0.00,-1.125)(1.406,-1.125)(0.00,0.00)
\qbezier(0.00,0.00)(-1.969,-0.375)(-1.125,0.750)
\qbezier(-1.125,0.750)(-0.281,1.875)(0.00,0.00)
\end{picture}}

\newcommand{\graphrrrb}{
\setlength{\unitlength}{0.8em}
\begin{picture}(3,2)(-0.25,-0.25)
\linethickness{0.1em}
\put(0.5,0){\circle*{.2}}
\linethickness{0.1em}
\put(2,0){\circle*{.2}}
\put(0.5,-0.5){\circle{1}}
\put(0.5,0.5){\circle{1}}
\put(2,0.5){\circle{1}}
\end{picture}
}

\newcommand{\graphrrrc}{
\setlength{\unitlength}{0.8em}
\begin{picture}(4.50,1.5)(-0.25,0)
\linethickness{0.1em}
\put(0.50,0.00){\circle*{.2}}
\put(0.50,0.50){\circle{1}}

\put(2,0.00){\circle*{.2}}
\put(2,0.50){\circle{1}}

\put(3.50,0.00){\circle*{.2}}
\put(3.50,0.50){\circle{1}}
\end{picture}
}

\newcommand{\graphgrra}{
\setlength{\unitlength}{0.8em}
\begin{picture}(2,2)(-0.25,-0.25)
\linethickness{0.1em}
\put(0.5,0){\circle*{.2}}
\put(1.5,0){\circle*{.2}}
\put(0.5,-0.5){\circle{1}}
\put(0.5,0.5){\circle{1}}

\linethickness{0.02em}
\put(0.5,0){\line(1,0){1}}
\end{picture}
}

\newcommand{\graphgrrb}{
\setlength{\unitlength}{0.8em}
\begin{picture}(3.5,2)(-0.25,-0.25)
\linethickness{0.1em}
\put(1,0){\circle*{.2}}
\put(2,0){\circle*{.2}}
\put(0.5,0){\circle{1}}
\put(2.5,0){\circle{1}}

\linethickness{0.05em}
\put(1,0){\line(1,0){1}}
\end{picture}
}

\newcommand{\graphgrrc}{
\setlength{\unitlength}{0.8em}
\begin{picture}(4,2)(-0.25,-0.25)
\linethickness{0.1em}
\put(1,0){\circle*{.2}}
\put(2,0){\circle*{.2}}
\put(3,0){\circle*{.2}}
\put(0.5,0){\circle{1}}
\put(3,0.5){\circle{1}}

\linethickness{0.05em}
\put(1,0){\line(1,0){1}}
\end{picture}
}

\newcommand{\graphgrrd}{
\setlength{\unitlength}{0.8em}
\begin{picture}(2.5,2)(-0.25,-0.25)
\linethickness{0.1em}
\put(1,0.5){\circle*{.2}}
\put(0.5,-0.5){\circle*{.2}}
\put(1.5,-0.5){\circle*{.2}}
\put(0.5,0.5){\circle{1}}
\put(1.5,0.5){\circle{1}}

\linethickness{0.05em}
\put(0.5,-0.5){\line(1,0){1}}
\end{picture}
}

\newcommand{\graphgrre}{
\setlength{\unitlength}{0.8em}
\begin{picture}(3,2)(-0.25,-0.25)
\linethickness{0.1em}
\put(0.5,0){\circle*{.2}}
\put(2,0){\circle*{.2}}
\put(0.75,-0.5){\circle*{.2}}
\put(1.75,-0.5){\circle*{.2}}
\put(0.5,0.5){\circle{1}}
\put(2,0.5){\circle{1}}

\linethickness{0.05em}
\put(0.75,-0.5){\line(1,0){1}}
\end{picture}
}

\newcommand{\graphggra}{
\setlength{\unitlength}{0.8em}
\begin{picture}(2.5,2)(-0.25,-0.25)
\linethickness{0.1em}
\put(0,0){\circle*{.2}}
\put(1,0){\circle*{.2}}
\put(0.5,0){\circle{1}}
\put(1.5,0){\circle{1}}
\end{picture}
}

\newcommand{\graphggrb}{
\setlength{\unitlength}{0.8em}
\begin{picture}(3,2)(-0.25,-0.25)
\linethickness{0.1em}
\put(0,0){\circle*{.2}}
\put(1,0){\circle*{.2}}
\put(2,-0.5){\circle*{.2}}
\put(0.5,0){\circle{1}}
\put(2,0){\circle{1}}
\end{picture}
}

\newcommand{\graphggrc}{
\setlength{\unitlength}{0.8em}
\begin{picture}(2.5,2)(-0.25,-0.25)
\linethickness{0.05em}
\put(1,-0.25){\line(1,0){1}}
\put(1,-0.25){\line(-1,0){1}}

\put(1,-0.25){\circle*{.2}}
\put(0,-0.25){\circle*{.2}}
\put(2,-0.25){\circle*{.2}}
\linethickness{0.1em}
\put(1,0.25){\circle{1}}
\end{picture}
}

\newcommand{\graphggrd}{
\setlength{\unitlength}{0.8em}
\begin{picture}(2.5,2)(-0.25,-0.25)
\linethickness{0.05em}
\put(0,0.75){\circle*{.2}}
\put(0,0.75){\line(0,-1){1}}
\put(0,-0.25){\circle*{.2}}
\put(0,-0.25){\line(1,0){1}}
\put(1,-0.25){\circle*{.2}}
\linethickness{0.1em}
\put(1.5,-0.25){\circle{1}}
\end{picture}
}

\newcommand{\graphggre}{
\setlength{\unitlength}{0.8em}
\begin{picture}(3,2)(-0.25,-0.25)
\linethickness{0.05em}
\put(0,0.5){\circle*{.2}}
\put(0,0.5){\line(0,-1){1}}
\put(0,-0.5){\circle*{.2}}
\put(0,-0.5){\line(1,0){1}}
\put(1,-0.5){\circle*{.2}}
\put(2,-0.5){\circle*{.2}}
\linethickness{0.1em}
\put(2,0){\circle{1}}
\end{picture}
}

\newcommand{\graphggrf}{
\setlength{\unitlength}{0.8em}
\begin{picture}(2.5,2)(-0.25,-0.25)
\linethickness{0.05em}
\put(0,0.5){\line(1,0){1}}
\put(0,-0.5){\line(1,0){1}}
\linethickness{0.05em}
\put(0,0.5){\circle*{0.2}}
\put(0,-0.5){\circle*{0.2}}
\put(1,0.5){\circle*{0.2}}
\put(1,-0.5){\circle*{0.2}}
\put(1.5,0.5){\circle{1}}
\end{picture}
}

\newcommand{\graphggrg}{
\setlength{\unitlength}{0.8em}
\begin{picture}(3,2)(-0.25,-0.25)
\linethickness{0.05em}
\put(0,0.5){\line(1,0){1}}
\put(0,-0.5){\line(1,0){1}}
\linethickness{0.05em}
\put(0,0.5){\circle*{0.2}}
\put(0,-0.5){\circle*{0.2}}
\put(1,0.5){\circle*{0.2}}
\put(1,-0.5){\circle*{0.2}}
\put(1.5,0){\circle*{0.2}}
\put(2,0){\circle{1}}
\end{picture}
}

\newcommand{\graphggga}{
\setlength{\unitlength}{0.8em}
\begin{picture}(1.5,2)(-0.25,-0.25)
\linethickness{0.05em}
\put(0,0){\circle*{.2}}
\put(0,0){\line(1,0){1}}
\put(1,0){\circle*{.2}}
\linethickness{0.1em}
\put(0.5,0){\circle{1}}
\end{picture}
}

\newcommand{\graphgggb}{
\setlength{\unitlength}{1em}
\begin{picture}(1.5,2)(-0.25,-0.25)
\linethickness{0.05em}
\put(0,-0.5){\circle*{.2}}
\put(0,-0.5){\line(1,1){1.0142}}
\put(0,-0.5){\line(1,0){1}}
\put(1,-.5){\circle*{.2}}
\put(1,-0.5){\line(0,1){1}}
\put(1,0.5){\circle*{.2}}
\end{picture}
}

\newcommand{\graphgggc}{
\setlength{\unitlength}{0.8em}
\begin{picture}(2.5,2)(-0.25,-0.25)
\linethickness{0.05em}
\put(0,0){\circle*{.2}}
\put(1,0){\circle*{.2}}
\put(1,0){\line(1,0){1}}
\put(2,0){\circle*{.2}}
\linethickness{0.1em}
\put(0.5,0){\circle{1}}
\end{picture}
}
\newcommand{\graphgggca}{
\setlength{\unitlength}{0.8em}
\begin{picture}(2.5,2)(-0.25,-0.25)
\linethickness{0.05em}
\put(0,0){\circle*{.2}}
\put(1,0){\circle*{.2}}
\put(1,0){\line(1,0){1}}
\put(2,0){\circle*{.2}}
\put(.5,0.375){\line(0,1){.25}}
\put(1.5,-0.125){\line(0,1){.25}}
\linethickness{0.1em}
\put(0.5,0){\circle{1}}
\end{picture}
}

\newcommand{\graphgggcb}{
\setlength{\unitlength}{0.8em}
\begin{picture}(2.5,2)(-0.25,-0.25)
\linethickness{0.05em}
\put(0,0){\circle*{.2}}
\put(1,0){\circle*{.2}}
\put(1,0){\line(1,0){1}}
\put(2,0){\circle*{.2}}
\put(.5,0.375){\line(0,1){.25}}
\put(.5,-0.375){\line(0,-1){.25}}
\linethickness{0.1em}
\put(0.5,0){\circle{1}}
\end{picture}
}

\newcommand{\graphgggd}{
\setlength{\unitlength}{0.8em}
\begin{picture}(2.5,2)(-0.25,-0.25)
\linethickness{0.05em}
\put(0,-0.5){\circle*{.2}}
\put(0,-0.5){\line(1,0){1}}
\put(1,-0.5){\circle*{.2}}
\put(1,-0.5){\line(1,0){1}}
\put(1,-0.5){\line(0,1){1}}
\put(1,0.5){\circle*{.2}}
\put(2,-0.5){\circle*{.2}}
\end{picture}
}

\newcommand{\graphggge}{
\setlength{\unitlength}{0.8em}
\begin{picture}(1.5,2)(-0.25,-0.25)
\linethickness{0.05em}
\put(0,0.5){\circle*{.2}}
\put(0,0.5){\line(0,-1){1}}
\put(0,-0.5){\circle*{.2}}
\put(0,-0.5){\line(1,0){1}}
\put(1,-0.5){\circle*{.2}}
\put(1,-0.5){\line(0,1){1}}
\put(1,0.5){\circle*{.2}}
\end{picture}
}

\newcommand{\graphgggf}{
\setlength{\unitlength}{0.8em}
\begin{picture}(1.5,2)(-0.25,-0.25)
\linethickness{0.05em}
\put(0,0.5){\circle*{.2}}
\put(0,-0.5){\circle*{.2}}
\put(1,0.5){\circle*{.2}}
\put(0,-0.5){\line(1,0){1}}
\put(1,-0.5){\circle*{.2}}
\linethickness{0.1em}
\put(0.5,0.5){\circle{1}}\end{picture}
}

\newcommand{\graphgggg}{
\setlength{\unitlength}{0.8em}
\begin{picture}(1.5,2)(-0.25,-0.25)
\linethickness{0.05em}
\put(0,0.75){\circle*{.2}}
\put(0,0.75){\line(1,0){1}}
\put(1,0.75){\circle*{.2}}
\put(1,0.75){\line(0,-1){1}}
\put(1,-0.25){\circle*{.2}}

\put(0,-0.75){\circle*{.2}}
\put(0,-0.75){\line(1,0){1}}
\put(1,-0.75){\circle*{.2}}
\end{picture}
}

\newcommand{\graphgggh}{
\setlength{\unitlength}{0.8em}
\begin{picture}(1.5,2)(-0.25,-0.25)
\linethickness{0.05em}
\put(0,0.5){\circle*{.2}}
\put(0,0.5){\line(1,0){1}}
\put(1,0.5){\circle*{.2}}

\put(0,0){\circle*{.2}}
\put(0,0){\line(1,0){1}}
\put(1,0){\circle*{.2}}

\put(0,-0.5){\circle*{.2}}
\put(0,-0.5){\line(1,0){1}}
\put(1,-0.5){\circle*{.2}}
\end{picture}
}

\newcommand{\labeledgraphv}[1]{
\setlength{\unitlength}{1em}
\begin{picture}(1,1)(0.8,0.25)
\linethickness{0.1em}
\put(1,.5){\circle*{.2}}
\put(1.5,0.25){\scriptsize{$#1$}}
\linethickness{0.1em}
\end{picture}
}

\newcommand{\labeledgraphr}[1]{
\setlength{\unitlength}{1em}
\begin{picture}(3,1)(-0.6,0.25)
\linethickness{0.1em}
\put(1,.5){\circle*{.2}}
\put(1.5,0.25){\scriptsize{$#1$}}
\linethickness{0.1em}
\put(0.5,0.5){\circle{1}}
\end{picture}
}

\newcommand{\labeledgraphgamma}[2]{
\setlength{\unitlength}{1em}
\begin{picture}(2,1)(-0.2,0)
\linethickness{0.1em}
\put(0.25,0){\circle*{.2}}
\put(0.25,0){${}^{#1}$}
\put(1.25,0){\circle*{.2}}
\put(1.25,0){${}^{#2}$}
\linethickness{0.05em}
\put(0.25,0){\line(1,0){1}}
\end{picture}
}

\newcommand{\labeledgraphgra}[2]{
\setlength{\unitlength}{1em}
\begin{picture}(2.5,2)(0,0)
\linethickness{0.1em}
\put(0.5,1){\circle*{.2}}
\put(-0.5,1.5){\scriptsize $#1$}
\put(1.5,1){\circle*{.2}}
\put(2,0.25){\scriptsize $#2$}
\put(0.5,0.5){\circle{1}}
\linethickness{0.05em}
\put(0.5,1){\line(1,0){1}}
\end{picture}
}

\newcommand{\td}[2]{\frac{d #1}{d #2}}

\newtheorem{definition}{Definition}

\newcommand{\stackleft}[2]{ \,{}^{(#1)}_{\ #2}}
\newcommand{\stacklr }[5]{ \,{}^{(#1)}_{\ #2}#3^{#4}_{#5}\,}
\newcommand{\sigmaone}[3]{\left[ {}^{(1)}_{1} \sigma_{#2
 }^#1\right]_{#3}}
\newcommand{\sigmatwo}[5]{\left[ {}^{(0)}_2 \sigma_{#2 #3}^#1\right]_{#4,#5}}
\newcommand{\sigmathree}[7]{\left[{}^{(1)}_3\sigma_{#2 #3 #4}^#1\right]_{#5,#6,#7}}

\newcommand{\tauone}[3]{\left[ {}^{(1)}_{1} \tau_{#2
 }^#1\right]_{#3}}

\newcommand{\tauthree}[7]{\left[{}^{(1)}_3\tau_{#2 #3 #4}^#1\right]_{#5,#6,#7}}

\newcommand{\betaoneG}[3]{\left[ {}^{(1)}_{1} \beta^{#1}\left(#2\right)\right]_{#3}}
\newcommand{\betatwoG}[5]{\left[ {}^{(0)}_{2} \beta^{#1 #2}\left(#3\right)\right]_{#4,#5}}
\newcommand{\betathreeG}[7]{\left[ {}^{(1)}_{3} \beta^{#1#2#3}\left(#4\right)\right]_{#5,#6,#7}}

\renewcommand{\stacklr}[5]{ \,{}^{(#1)}_{\ #2}#3^{#4}_{#5}\,}

\newcommand{\stacklrsql}[5]{ {}^{(#1)}_{\ #2}#3^{#4}_{#5}\,}
\newcommand{\stacklrcncn}[5]{ {}^{(#1)}_{\ #2}#3^{#4}_{#5}\,}

\newcommand{\stacklru}[5]{ \,{}^{(#1)}_{\ #2}{#3_{#4}}^{#5}\,}
\newcommand{\stacklrl}[5]{ \,{}^{(#1)}_{\ #2}#3^{#4}_{#5}\,}

\newcommand{\Rmn}[1]{\uppercase\expandafter{\romannumeral #1}}
\usepackage{rotating}
\begin{document}

\title{On the Use of Group Theoretical and Graphical Techniques
toward the Solution of the General $\bm{N}$-body Problem}

\bigskip
\bigskip
\author{W.B.\ Laing\footnote{Current Address: Department of Physics, Kansas State University}, M.\ Dunn, D.K.\ Watson}
\affiliation{Homer L. Dodge Department of Physics and Astronomy, University of Oklahoma, Norman, OK 73019}
\date{\today}

\begin{abstract}

Group theoretic and graphical techniques are used to derive the $N$-body
wave function for a system of identical bosons with general interactions 
through first-order in a perturbation
approach.  This method is based on the maximal symmetry present at lowest order
in a perturbation series in inverse spatial dimensions. The symmetric
structure at lowest order has a point group isomorphic with the $S_{N}$
group, the symmetric group of $N$ particles, and the resulting perturbation
expansion of the Hamiltonian is
order-by-order invariant under the permutations of the $S_{N}$ group.  
This invariance under $S_N$
imposes severe
symmetry requirements on the tensor blocks needed at each order in
the perturbation
series. We show here that these blocks can be decomposed into
a basis of binary tensors invariant under $S_{N}$. This basis is small
(25 terms at first order in the wave function), independent of $N$, and is
derived using graphical techniques.
This checks the $N^{6}$ scaling of these terms
at first order by effectively separating the $N$ scaling problem away from
the rest of the physics. 
The transformation of each binary tensor to the final normal coordinate basis
requires the derivation of Clebsch-Gordon coefficients of $S_N$ for arbitrary
$N$. This has been accomplished using the group theory
of the symmetric group.  
This achievement results in an analytic solution
for the wave function,
exact through first order, that scales as $N^0$\,, effectively
circumventing intensive numerical work. This solution can be
systematically improved with further analytic work by going to yet higher orders in the perturbation series.\end{abstract}
\maketitle

\section{Introduction}

Obtaining solutions for large systems of interacting particles continues to
challenge existing approaches and current numerical resources. As the
number of particles $N$ increases, the Hilbert space that holds an exact
solution of the problem scales exponentially in $N$ making
a direct numerical simulation intractable.\cite{liu:2007,montina2008}  
For general interparticle
interactions, this necessitates approximations which in general truncate
the Hilbert space of the solution, usually by choosing a
particular ansatz for the manybody wave function  or by truncating a
perturbation series. Successful methods include
mean-field theory and its variants\cite{dalfovo1999}, the self-consistent
multiconfigurational Hartree Fock theory\cite{masiello2005},
coupled cluster methods which are size consistent\cite{CC},
Monte Carlo methods\cite{Reynolds:82,Hommond:94,Bressanini:99,dubois:01,dubois:03,purwanto:05,mc, MinguzziAnderson},  and
density functional methods which avoid an explicit determination of the
$N$-body wave function.\cite{nunes1999,singh}
Perturbation techniques have  typically been based on truncated
expansions in the interaction strength and thus have been successful in
a limited range of the interaction strength.\cite{dalfovo1999}

In the current paper, we address the challenge of
the exponential $N$ scaling for systems of $N$ identical bosons by using
group theory and by employing
graphical techniques.
This approach is thus inherently analytical in nature although the potential
is general.  It is based
on the maximal symmetry present at lowest order in a perturbation series
in the inverse dimension
of space. The lowest-order configuration has a point group isomorphic with
the symmetric group, $S_N$\cite{hamermesh}, with all interparticle distances and angles
identical (achievable only in higher dimensions). This simple
configuration at lowest order yields to an closed form analytic solution  which can be shown
to contain correlation effects.

The coefficients needed at each order in the perturbation series, can be grouped into tensors in the particle label space. These coefficient tensors 
in the higher-order terms
scale as $(N^2)^{(Order+2)}$ ($N^6$ for first order in the wave 
function) reflecting the $N$ dependence of the size of the initial internal
coordinate basis. However, due to the expansion about the large-dimension, maximally-symmetric structure in which all interparticle distances are equal,
the size of the internal coordinate basis does not determine the $N$ scaling
of the problem.
As explained in more detail in the body of the paper,
the system is order-by-order invariant under the $N!$ operations of the $S_N$ point group which 
imposes a highly symmetric
structure on the tensor blocks needed at each order. We show here that
these blocks can be
systematically decomposed using a basis of binary tensors invariant under
$S_N$ which are derived in this paper using graphical techniques.
This basis is finite,
independent of $N$, and in fact small (only twenty five terms at first 
order in the wave
function). Since the basis is small, there are only a small number of undetermined
coefficients in the wave function (twenty five at first order) for any $N$\,.
These few coefficients are determined by solving the perturbation equations.

Thus by expanding about a large dimension limit which provides the maximally
symmetric point group symmetry, we have separated the $N$ scaling problem away from the interaction dynamics. Then by allowing the $S_N$ symmetry to do the
"heavy lifting", we arrive at an approach to the $N$-body problem which scales
as $N^0$\,.
The number of particles
$N$ enters into the theory as a parameter so that results for a
broad range of $N$ can be obtained in a single analytic 
calculation.\cite{FGpaper,energy,paperI,zerothDensity}  
Choosing the
perturbation to be in the inverse spatial dimension, rather than in the
interaction strength,
allows this perturbation series to be used for both weakly and
strongly interacting series as well as the transition between them.



The final coordinate basis of the problem in which the perturbation
equations are solved, is a normal mode basis obtained 
from the
zeroth-order perturbation equation which is a harmonic equation (a brief four page
summary of the method at lowest order, along with some lowest-order results, may
be found in Ref.~\onlinecite{Laing:arXiv:physics/0510177v1}). The normal
modes transform under irreducible representations of the $S_N$ group and are
obtained using the FG method familiar from quantum chemistry.\cite{dcw}
They offer insight into
the dominant motions of the $N$-body system if higher order terms are small.
The transformation to normal modes 
transforms the basis of binary invariants into a basis of 
Clebsch-Gordon coefficients
of the symmetric group, $S_N$, which couple the different irreducible
representations together to form the scalar representation of $S_N$. This
transformation, as reported in this paper, has been achieved analytically
for arbitrary $N$ through first order (see Appendices B and C).

The final result is a solution for the wave function
of a system of $N$ identical particles,
interacting with a general
potential and confined in an external potential.
Such confined quantum systems are
now widespread across many areas of physics and possess from a few
tens to many millions of particles.
Quantum dots~\cite{quantumdots}, Bose-Einstein condensates~\cite{LegPit}, atoms, two-dimensional
electronic systems in a Corbino disk geometry~\cite{corbino} and rotating
superfluid helium systems~\cite{rsfhs} are all examples of quantum systems
confined in an external potential.
The interparticle interactions in these systems span an enormous
range of interaction strengths.
The solution involves a large number normal modes, however
the degeneracy of the corresponding frequencies for a system of $N$
identical bosons is extensive such that there are only five distinct 
frequencies for a spherical confining potential. Our
previous analysis shows that these correspond to:
breathing and center of mass motions (for a harmonic confining
potential), single particle excitations (two distinct frequencies)
and phonon modes.\cite{excitations} These normal modes will describe the dominant motions of
the above-mentioned manybody systems if higher-order terms are small.
The first-order correction to the Hamiltonian has terms cubic in the
normal coordinates and their derivatives.  This yields a first-order, anharmonic correction to the wave function that is equal to the lowest-order state times a
third-order polynomial of odd powers in the normal modes.

The approach developed in this paper is generalizable to higher order. This
is discussed in the Summary and Conclusions.

The structure of this paper is as follows. 
In Section~\ref{sec:DPT}, we set up the formalism
for the perturbation series in the inverse dimension of space through first
order in the wave function. Expansions in the inverse dimension of space
have been developed for many different systems in physics.\cite{generaldpt}
The approach used
in this paper\cite{copen92,chattrev}
is outlined for a general $N$-particle problem for $L=0$ in this section.
In Section~\ref{sec:PGSI}, we discuss the maximal point group symmetry of the
zeroth order ($D \longrightarrow \infty$)
system and its implications. The presence of this extremely high degree of
symmetry makes possible
the solution of the perturbation equations for such a complex, and possibly
strongly interacting, system.
In Section~\ref{sec:NMHWF}, we review the solution for the zeroth-order
wave function
in preparation for the derivation of the first-order, anharmonic correction.
In Section~\ref{sec:FAWF}, we develop the mathematical formalism
needed to solve for the first-order wave function correction and, as an
example, apply it to deriving the first-order
wave function correction for the ground state of a system
of $N$ identical bosons. In Section~\ref{sec:SumCon}, we present a summary and our conclusions.

The formalism developed in this paper, and briefly outlined above, has been 
tested on the example of a three-dimensional fully-interacting confined 
system of $N$ particles.\cite{wavefunction1harm}
The interaction and confining potential were both chosen to be harmonic.
Choosing this simple interaction is not necessary or even advantageous
for our method, but it did allow an independent exact solution    of a 
fully-interacting three-dimensional system for 
comparison with the results of our formalism. This was achieved by 
comparing the general wave function through first-order of 
Eqs.~(\ref{eq:Phit1}) and (\ref{eq:deltacommute})  with the
corresponding first-order wave function derived by expanding the independent
solution to first order. The two, 
independently derived, first-order wave functions are seen to be the 
same for any $N$\,.

In a second example\cite{density1harm} we show how properties may be 
derived from the wave function of 
Eqs.~(\ref{eq:Phit1}) and (\ref{eq:deltacommute}), by deriving
a closed-form expression for the 
density profile, an observable measurable in the laboratory
for systems such as a Bose-Einstein condensate. 
This general result for the density profile is checked by comparing with 
an independent solution of the above mentioned system, $N$ particles interacting
harmonically in a harmonic confining potential. The first-order density profile
from the present
formalism is indistinguishable from the independent solution and shows strong
convergence to the exact result to all orders.

\section{The Perturbation Formalism in Inverse Spatial Dimensions}
\label{sec:DPT}
\subsection{The Dimensional Expansion: Dimensionally-Scaled,
  Jacobian-Weighted $S$-wave Hamiltonian in Internal Coordinates}
In previous papers we have started the development of
the dimensional expansion formalism for the quantum mechanical problem of $N$ interacting bosons
confined by an external potential, where $N$ may be less than
ten to $10^7$ or larger. The zeroth-order wave function, density
profile and corresponding energy have been previously
derived.\cite{loeser,FGpaper,energy,paperI,zerothDensity}

In dimensionally-scaled oscillator units the Jacobian-weighted, $L=0$
Schr\"{o}dinger equation reads\cite{avery,paperI}
\begin{equation} \label{eq:scale1}
\bar{H} \Phi =
\left(\frac{1}{\kappa(D)}\bar{T}+\bar{V}_{\mathtt{eff}}
\right)\Phi = \bar{E} \Phi,
\end{equation}
where
\begin{widetext}
\begin{equation}
\bar{T} 
= 
{\displaystyle \hbar^2 \Biggl[-\sum\limits_{i=1}^{N}\frac{1}{2
m_i} \frac{\partial^2}{{\partial \bar{r}_i}^2}-
\sum\limits_{i=1}^{N}\frac{1}{2 m_i \bar{r}_i^2}
\sum\limits_{j\neq i}^N \sum\limits_{k\neq i}^N
\frac{\partial}{\partial\gamma_{ij}}
(\gamma_{jk}-\gamma_{ij} \gamma_{ik})
\frac{\partial}{\partial\gamma_{ik}}
 \Biggr] } \,.
\label{eq:SE_T}
\end{equation}
\end{widetext}
The effective potential $\bar{V}_{\mathtt{eff}}$ is composed of the non-derivative (``centrifugal'') portion of the kinetic
energy $\bar{U}$\,, plus the confining and interparticle potentials, $\bar{V}_{\mathtt{conf}}$ and $\bar{V}$ respectively, i.e.\
\begin{equation}
\bar{V}_{\mathtt{eff}} = \bar{U} + \bar{V}_{\mathtt{conf}}+\bar{V} \,,
\end{equation}
where
\begin{equation}
\bar{V}_{\mathtt{conf}} = \sum_{i=1}^N \bar{v}_{\mathtt{conf}}(\bar{r}_i)
\end{equation}
\begin{equation}
\bar{U} 
= 
{\displaystyle \sum\limits_{i=1}^{N} 
\hbar^2
\frac{N(N-2)+(D-N-1)^2 \left( \frac{\Gamma^{(i)}}{\Gamma} \right)
}{\kappa(D)\, 8 m_i \bar{r}_i^2} 
 }
\end{equation}
\begin{equation}
\bar{V} = \sum\limits_{i=1}^{N-1}\sum\limits_{j=i+1}^{N}
\bar{V}_{\mathtt{int}}(\bar{r}_{ij}) \,.
\end{equation}
The $\lim_{D \rightarrow \infty} \kappa(D) \propto D^2$ to
ensure that $\bar{U}$ is finite as $D \rightarrow \infty$\,. Defining $\delta$ to be the inverse dimensionality (so that $\lim_{D \rightarrow \infty}$ is equivalent to $\lim_{\delta \rightarrow 0}$) and $\zeta(\delta)$ to be the ratio of $D^2$ to $\kappa(D)$,
\begin{eqnarray}
 \delta&=&\frac{1}{D}
 \\
 \zeta(\delta)&=&\frac{1}{\delta^2\kappa(D)} \,, \label{eq:kappa_exp}
\end{eqnarray}
then $ \zeta(\delta)$ is finite as $\delta \longrightarrow 0$ ($D \longrightarrow \infty$).
Defining a position vector, ${\bar{\bm{y}}}$\,, consisting of all
$N(N+1)/2$ internal coordinates:
\begin{equation}\label{eq:ytranspose}
\begin{array}[t]{l} {\bar{\bm{y}}} = \left( \begin{array}{c} \bar{\bm{r}} \\
\bm{\gamma} \end{array} \right) \,, \;\;\; \mbox{where} \;\;\; \\
\mbox{and} \;\;\; \bar{\bm{r}} = \left(
\begin{array}{c}
\bar{r}_1 \\
\bar{r}_2 \\
\vdots \\
\bar{r}_N
\end{array}
\right) \,. \end{array} 
\bm{\gamma} = \left(
\begin{array}{c}
\gamma_{12} \\ \cline{1-1}
\gamma_{13} \\
\gamma_{23} \\ \cline{1-1}
\gamma_{14} \\
\gamma_{24} \\
\gamma_{34} \\ \cline{1-1}
\gamma_{15} \\
\gamma_{25} \\
\vdots \\
\gamma_{N-2,N} \\
\gamma_{N-1,N} \end{array} \right) \,,
\end{equation}
and taking all of the masses of the particles to be the same, $m_i
= m$\,, Eq.~(\ref{eq:scale1}) can be written as
\begin{equation} \label{eq:scale2}
\bar{H} \Phi = \left(\delta^2 \,\, \zeta(\delta) \,\bar{T}+\bar{V}_{\mathtt{eff}}
\right)\Phi = \bar{E} \Phi,
\end{equation}
where
\begin{widetext}
\begin{equation}
 \zeta(\delta) \, \bar{T} 
= 
{\displaystyle - \frac{1}{2} \partial_{{\bar{y}}_{\nu_1}} \,
G({\bar{\bm{y}}}; \delta)_{\nu_1,\nu_2} \,
\partial_{{\bar{y}}_{\nu_2} }} \,, \label{eq:SE_T2}
\end{equation}
\end{widetext}
%
\begin{equation}
\bar{V}_{\mathtt{eff}}({\bar{\bm{y}}}; \delta, N) =
\bar{U}({\bar{\bm{y}}}; \delta, N) + \bar{V}_{\mathtt{conf}}({\bar{\bm{y}}}; \delta, N) + \bar{V}({\bar{\bm{y}}};
\delta, N) \,,
\end{equation}
\begin{equation}
\bar{U}({\bar{\bm{y}}}; \delta, N) 
= 
{\displaystyle \sum\limits_{i=1}^{N} 
\hbar^2 \zeta(\delta)
\frac{N(N-2) \delta^2 +(1-(N+1) \delta)^2 \left(
\frac{\Gamma^{(i)}}{\Gamma} \right) }{ 8 m \bar{r}_i^2}  
} \,,
\end{equation}
$\partial_{{\bar{y}}_{\nu}} = \frac{\partial}{\partial
{\bar{y}}_\nu}$ are the derivatives of the elements of the
${\bar{\bm{y}}}$ column vector, and $\bm{G}({\bar{\bm{y}}}; \delta)$
is the $N(N+1)/2 \times N(N+1)/2$ dimensional block-diagonal tensor
\begin{equation}
\bm{G}({\bar{\bm{y}}}; \delta) = \left( \begin{array}{c|c}
{\displaystyle \frac{\hbar^2 \, \zeta(\delta)}{m} \bm{I}_N } & \bm{0} \\
\hline \bm{0} & \bm{G}^{\gamma \gamma}(\bar{\bm{r}}, \bm{\gamma};
\delta)
\end{array} \right)
\end{equation}
with $\bm{I}_N$ the $N \times N$ dimensional diagonal unit tensor,
$\bm{G}^{\gamma \gamma}(\bar{\bm{r}}, \bm{\gamma}; \delta)$ the
$N(N-1)/2 \times N(N-1)/2$ dimensional matrix of elements
\begin{equation} \label{eq:gmfgg}
[\bm{G}^{\gamma \gamma}(\bar{\bm{r}}, \bm{\gamma};
\delta)]_{(ij),(lk)} = \frac{\hbar^2 \, \zeta(\delta)}{m} \,
\frac{(\gamma_{jk}-\gamma_{ij} \gamma_{lk}) }{4 \, \bar{r}_i^2} \,\,
\delta_{il} \,,
\end{equation}
and $\delta_{il}$ is the Kronecker delta (as distinct from the
inverse spatial dimension, $\delta$). This paper uses the implicit summation
convention, i.e.\ repeated indices are summed over (unless explicitly noted otherwise), and this
convention has been used in Eq.~(\ref{eq:SE_T2}).

\subsection{The Large-Dimension Expansion}
\subsubsection{The Large-Dimension Limit}
\label{subsubsec:LDL}
Looking at Eq.~(\ref{eq:scale2}), we see that as $\delta
\rightarrow 0$ ($D \rightarrow \infty$) the derivative terms of
Eq.~(\ref{eq:SE_T2}) drop out leaving a static problem where the
system localizes at the minima (or more generally the extrema) of
$\bar{V}_{\mathtt{eff}}({\bar{\bm{y}}}; \delta=0, N)$ at
$\bar{r}_{i} = \bar{r}_{\infty}$ and $\gamma_{ij} =
\overline{\gamma}_{\infty}$\,. Thus we have the $D \rightarrow
\infty$ energy $\bar{E}_{\infty} = \bar{H}_{\infty} =
\left[\bar{V}_{\mathtt{eff}} \right]_{\delta^{1/2}=0}$ \,.

\subsubsection{Dimensionally-Scaled Internal Displacement Coordinates}
The dimensional expansion is developed by making the
following substitutions for all radii and angle cosines:
\begin{equation} \label{eq:taylor1}
\bar{r}_{i} = \bar{r}_{\infty}+\delta^{1/2}\,\bar{r}'_{i} \;\;\;
\mbox{and} \;\;\; \gamma_{ij} =
\overline{\gamma}_{\infty}+\delta^{1/2}\,\overline{\gamma}'_{ij} \,,
\end{equation}
and defining a displacement vector of the internal displacement
coordinates [primed in Eqs.~(\ref{eq:taylor1})]
\begin{equation}\label{eq:ytransposeP}
\begin{array}[t]{l} {\bar{\bm{y}}'} = \left( \begin{array}{c} r \\
\overline{\bm{\gamma}}' \end{array} \right) \,, \;\;\;
\mbox{where} \;\;\; \\ \mbox{and} \;\;\; \bar{\bm{r}}' = \left(
\begin{array}{c}
\bar{r}'_1 \\
\bar{r}'_2 \\
\vdots \\
\bar{r}'_N
\end{array}
\right) \,, \end{array} 
\overline{\bm{\gamma}}' = \left(
\begin{array}{c}
\overline{\gamma}'_{12} \\ \cline{1-1}
\overline{\gamma}'_{13} \\
\overline{\gamma}'_{23} \\ \cline{1-1}
\overline{\gamma}'_{14} \\
\overline{\gamma}'_{24} \\
\overline{\gamma}'_{34} \\ \cline{1-1}
\overline{\gamma}'_{15} \\
\overline{\gamma}'_{25} \\
\vdots \\
\overline{\gamma}'_{N-2,N} \\
\overline{\gamma}'_{N-1,N} \end{array} \right) \,.
\end{equation}

\subsubsection{Dimensional Perturbation Series}
After the substitutions of Eqs.~(\ref{eq:taylor1}) and (\ref{eq:ytransposeP}) in Eqs.~(\ref{eq:SE_T2})-(\ref{eq:gmfgg}), the Hamiltonian and Jacobian-weighted wave function and
energy are expanded in powers of $\delta^{1/2}$ by deriving Taylor
expansions in $\delta^{1/2}$ for
$\bm{G}\left({\bar{\bm{y}}}(\delta^{1/2}); \delta \right)$ and
$\bar{V}_{\mathtt{eff}}\left({\bar{\bm{y}}}(\delta^{1/2}); \delta,
N\right)$ to obtain
\begin{equation} \label{eq:dpt_exp}
\renewcommand{\arraystretch}{2} \begin{array}{r@{}l@{}c}
{\displaystyle \bar{H} = \bar{H}_{\infty} + \delta^{\frac{1}{2}}
\, \bar{H}_{-1} + \delta} & {\displaystyle \, \sum_{j=0}^\infty
\left(\delta^{\frac{1}{2}}\right)^j 
\bar{H}_j } & \\
{\displaystyle \Phi(\bar{r}_i,\gamma_{ij}) = } & {\displaystyle \,
\sum_{j=0}^\infty \left(\delta^{\frac{1}{2}}\right)^j 
\Phi_j }& \\
{\displaystyle \bar{E} = \bar{E}_{\infty} + \delta^{\frac{1}{2}}
\, \bar{E}_{-1} + \delta} & {\displaystyle \, \sum_{j=0}^\infty
\left(\delta^{\frac{1}{2}}\right)^j 
\bar{E}_j } &
 \,.
\end{array}
\renewcommand{\arraystretch}{1}
\end{equation}

The expansion of $\bm{G}\left({\bar{\bm{y}}}(\delta^{1/2}); \delta
\right)$ and
$\bar{V}_{\mathtt{eff}}\left({\bar{\bm{y}}}(\delta^{1/2}); \delta,
N\right))$ is most easily achieved using the identities of
Appendix~\ref{app:identities}. These identities show that to a given
order $j$ in $\delta^{1/2}$\,, the expansion terms are polynomials of
order $j$ of the elements of the ${\bar{\bm{y}}'}$ vector,
${\bar{\bm{y}}\prime}_\nu$\,. If $\bm{G}({\bar{\bm{y}}}; \delta)$
and $\bar{V}_{\mathtt{eff}}({\bar{\bm{y}}}; \delta, N)$ are
functions of $\delta$\,, rather than $\delta^{1/2}$\,, then the
equations of Appendix ~\ref{app:identities} indicate that these
polynomials of order $j$ are either all even or all odd in powers of
${\bar{\bm{y}}\prime}_\nu$ depending on whether $j$ is even or odd.

Thus we have that\cite{covcont}
\begin{eqnarray}
\bar{H}_{-1} & = & {\bar{\bm{y}}^\prime}_\nu \,\,
\partial_{{\bar{y}}_{\nu}} \,
\bar{V}_{\mathtt{eff}}({\bar{\bm{y}}}; \delta = 0,
N)|_{{\bar{\bm{y}}} = {\bar{\bm{y}}}_\infty} = 0\,, \\
\bar{H}_{0} & = & -\frac{1}{2}\stacklrl{0}{2}{G}{}{\nu_1,\nu_2}
\partial_{{\bar{y}^\prime}_{\nu_1}}
\partial_{{\bar{y}^\prime}_{\nu_2 }} +
\frac{1}{2}\stacklrl{0}{2}{F}{}{\nu_1,\nu_2}
\bar{y}^\prime_{\nu_1 } \label{eq:harm_H}
\bar{y}^\prime_{\nu_2}+\stacklrl{0}{0}{F}{}{} \,, \label{eq:H0y} \\
\bar{H}_{1} & = &-\frac{1}{2}\stacklrl{1}{3}{G}{}{\nu_1,\nu_2,\nu_3}
\bar{y}^\prime_{\nu_1 } \partial_{{\bar{y}^\prime}_{\nu_2 }}
\partial_{{\bar{y}^\prime}_{\nu_3 }}
-\frac{1}{2}\stacklrl{1}{1}{G}{}{\nu}
\partial_{{\bar{y}^\prime}_\nu}
+\frac{1}{3!}\stacklrl{1}{3}{F}{}{\nu_1,\nu_2,\nu_3}
\bar{y}^\prime_{\nu_1 } \bar{y}^\prime_{\nu_2}
\bar{y}^\prime_{\nu_3 } +\stacklrl{1}{1}{F}{}{\nu}
\bar{y}^\prime_\nu \,, \label{eq:one_H}
\end{eqnarray}
where the superprescript on the $\bm{F}$ and $\bm{G}$ tensors in
parentheses denotes the order in $\delta^{1/2}$ that the term enters
(harmonic being zeroth order); the subprescripts denote the rank of
the tensors, and
\begin{eqnarray}
\stacklrl{0}{2}{G}{}{\nu_1,\nu_2} & = & \bm{G}({\bar{\bm{y}}} =
{\bar{\bm{y}}}_\infty;
\delta=0)_{\nu_1,\nu_2} \,, 
\\
\stacklrl{1}{3}{G}{}{\nu_1,\nu_2,\nu_3} & = & \left.
\partial_{{\bar{y}}_{\nu_1}} \bm{G}({\bar{\bm{y}}};
\delta=0)_{\nu_2,\nu_3} \, \right|_{{\bar{\bm{y}}} = {\bar{\bm{y}}}_\infty} \,, \\
\stacklrl{1}{1}{G}{}{\nu_1} & = & \stacklrl{1}{3}{G}{}{\nu,\nu,\nu_1}
\,, 
\\
\stacklrl{0}{2}{F}{}{\nu_1,\nu_2} & = &
\partial_{{\bar{y}}_{\nu_1}} \partial_{{\bar{y}}_{\nu_2}}
\bar{V}_{\mathtt{eff}}({\bar{\bm{y}}}; \delta = 0,
N)|_{{\bar{\bm{y}}} = {\bar{\bm{y}}}_\infty} \,,
\\
\stacklrl{0}{0}{F}{}{} & = & \left.\frac{d}{d \delta} \,
\bar{V}_{\mathtt{eff}}({\bar{\bm{y}}}  =
{\bar{\bm{y}}}_\infty; \delta, N)\right|_{\delta=0} \,, 
\\
\stacklrl{1}{3}{F}{}{\nu_1,\nu_2,\nu_3} & = &
\partial_{{\bar{y}}_{\nu_1}} \partial_{{\bar{y}}_{\nu_2}}
\partial_{{\bar{y}}_{\nu_3}} \bar{V}_{\mathtt{eff}}({\bar{\bm{y}}};
\delta = 0,
N)|_{{\bar{\bm{y}}} = {\bar{\bm{y}}}_\infty} \,,
\\
\stacklrl{1}{1}{F}{}{\nu} & = & \left.
\partial_{{\bar{y}}_{\nu}} \left(\left.\frac{d}{d \delta}
\bar{V}_{\mathtt{eff}}({\bar{\bm{y}}}; \delta,
N)\right|_{\delta=0}\right) \right|_{{\bar{\bm{y}}} =
{\bar{\bm{y}}}_\infty} \,.
\end{eqnarray}

\subsection{The Solution of the Perturbation Equations}
\subsubsection{The Lowest-Order Wave Function} \label{para:wavef}
The zeroth-order Hamiltonian of Eq.~(\ref{eq:H0y}) is that of a multi-dimensional
harmonic oscillator. Thus upon transformation to the normal
modes of the system, the wave function is a product of $P = N(N+1)/2$ one-dimensional harmonic
oscillator wave functions
\begin{equation}
\Phi_0({\mathbf{\bar{y}'}}) = \prod_{\nu=1}^P
\phi_{n_{\nu}}\left(  \sqrt{\bar{\omega}_\nu} 
\, q^\prime_{\nu} \right) \,, \label{eq:psiintro}
\end{equation}
where $\bar{\omega}_\nu$ is the frequency of normal mode $q^\prime_{\nu}$\,,
and $n_{q^\prime_{\nu}}$ is the oscillator quantum number, $0 \leq n_{q^\prime_{\nu}}
< \infty$, which counts the number of quanta in this normal mode.

The transformation to normal modes would appear to be a formidable proposition since if $N$ is in the millions,
the number of normal modes $P$ is of the order $10^{12}$ or larger. However, the $D \rightarrow \infty$ structure is extremely symmetric (see Section~\ref{subsec:PGS}),  and it is this maximal point group symmetry which allows the normal modes to be derived (see Sec.~\ref{subsec:LowestOrder}).

\subsubsection{Next-Order, Beyond-Harmonic Energy}
\label{para:e1}
The first-order, beyond-harmonic energy correction is
\begin{equation}
\bar{E}_{1} = \hspace{0.5ex} \langle \Phi_{0} | \bar{H}_{1} | \Phi_0 \rangle \,.
\end{equation}

This is equal to zero which can be seen as follows. The harmonic-order wave function, $\Phi_0$\,, is a separable product of harmonic
oscillator functions. When $\bar{H}_{1}^{_{}}$ is written in terms
of normal coordinates, it is composed of odd powers of
normal coordinates and their derivatives (see Eq. (\ref{eq:H1V})). Thus if
$\phi_{n_{\nu}}\left(  \sqrt{\bar{\omega}_\nu} 
\, q^\prime_{\nu} \right)$ is a one-dimensional harmonic oscillator function
\begin{equation}
  \int^{+ \infty}_{- \infty} \phi_{n_{\nu}}\left(  \sqrt{\bar{\omega}_\nu} 
\, q^\prime_{\nu} \right) q_{\nu}^{\prime a} \partial^{b}_{q^\prime_{\nu}} \phi_{n_{\nu}}\left(  \sqrt{\bar{\omega}_\nu} 
\, q^\prime_{\nu} \right) dq^\prime_{\nu}= 0
\end{equation}
when $a + b = odd$, i.e. we have that
\begin{equation}
  \bar{E}_{1} = 0 \,.
\end{equation}

\subsubsection{Next-Order, Beyond-Harmonic Wave Function}
While the $O(\delta^{3 / 2})$ energy correction is zero, the
corresponding wave function correction is NOT zero. Thus if we write the Jacobian-weighted wave function through
first-order beyond harmonic as
\begin{equation}
\label{eq:Phit1}
\Phi = (1 + \delta^{1 / 2} \Delta_{1}) \Phi_{0} + O (\delta) \,,
\end{equation}
then $\Delta_{1}$ satisfies the commutator equation
\begin{equation} \label{eq:deltacommute}
[\Delta_{1}\,, \, \bar{H}_{0}] \, \Phi_{0} = \bar{H}_{1}
\Phi_0 \,.
\end{equation}

\section{The Maximal Point Group Symmetry and Its Implications}
\label{sec:PGSI}
\subsection{The Block Structure of the $\bm{F}$ and $\bm{G}$ Tensors}
If we generically denote the $\bm{F}$ and $\bm{G}$ tensors by
$\bm{Q}$ then we can write the $\stacklrl{0}{2}{\bm{Q}}{}{}$ tensors
in block form as
\begin{equation}\label{eq:Qblocks2}
\renewcommand{\arraystretch}{1.5}
\stacklr{0}{2}{\bm{Q}}{}{}=\left(
\begin{array}{ll}
\stacklr{0}{2}{\bm{Q}}{rr}{} & \stacklr{0}{2}{\bm{Q}}{r\gamma}{} \\
\stacklr{0}{2}{\bm{Q}}{\gamma r}{} &
\stacklr{0}{2}{\bm{Q}}{\gamma\gamma}{}
\end{array}\right) \,.
\renewcommand{\arraystretch}{1}
\end{equation}
For example for the $\stacklr{0}{2}{\bm{G}}{}{}$ tensor this reads
\begin{equation} 
\renewcommand{\arraystretch}{1.5}
\stacklrcncn{0}{2}{G}{}{\nu_1,\nu_2}=\left(
\begin{array}{ll}
\stacklrcncn{0}{2}{G}{rr}{i,j} & \stacklrcncn{0}{2}{G}{r\gamma}{i,(jk)} \\
\stacklrcncn{0}{2}{G}{\gamma r}{(ij),k} &
\stacklrcncn{0}{2}{G}{\gamma\gamma}{(ij),(kl)}
\end{array}\right) \,.
\renewcommand{\arraystretch}{1}
\end{equation}
Likewise the $\stacklrl{1}{3}{\bm{Q}}{}{}$ tensor may be written in
block form as
\begin{equation} \label{eq:Qblocks3}
\renewcommand{\arraystretch}{1.5}
\stacklr{1}{3}{\bm{Q}}{}{}=\left(
\begin{array}{ll} \left( \begin{array}{l}
\stacklr{1}{3}{\bm{Q}}{rrr}{} \\
\stacklr{1}{3}{\bm{Q}}{r\gamma r}{} \end{array}
 \right) &
\left( \begin{array}{l} \stacklr{1}{3}{\bm{Q}}{rr\gamma}{} \\
\stacklr{1}{3}{\bm{Q}}{r\gamma\gamma}{} \end{array}
\right) \\[4ex]
\left( \begin{array}{l} \stacklr{1}{3}{\bm{Q}}{\gamma rr}{} \\
\stacklr{1}{3}{\bm{Q}}{\gamma \gamma r}{} \end{array}
 \right) & \left(
\begin{array}{l} \stacklr{1}{3}{\bm{Q}}{\gamma r \gamma
}{}\\\stacklr{1}{3}{\bm{Q}}{\gamma\gamma\gamma}{}
\end{array}  \right)
\end{array}\right) \,,
\renewcommand{\arraystretch}{1}
\end{equation}
where, for example, the $\stacklrl{1}{3}{\bm{F}}{}{}$ tensor may be
written in block form as
\begin{equation} 
\renewcommand{\arraystretch}{1.5}
\stacklrsql{1}{3}{F}{}{\nu_1,\nu_2,\nu_3}=\left(
\begin{array}{ll} \left( \begin{array}{l}
\stacklrsql{1}{3}{F}{rrr}{i,j,k} \\
\stacklrsql{1}{3}{F}{r\gamma r}{i,(jk),l} \end{array}
 \right) &
\left( \begin{array}{l} \stacklrsql{1}{3}{F}{rr\gamma}{i,j,(kl)} \\
\stacklrsql{1}{3}{F}{r\gamma\gamma}{i,(jk),(lm)} \end{array}
\right) \\[4ex]
\left( \begin{array}{l} \stacklrsql{1}{3}{F}{\gamma rr}{(ij),k,l} \\
\stacklrsql{1}{3}{F}{\gamma \gamma r}{(ij),(kl),m} \end{array}
 \right) & \left(
\begin{array}{l} \stacklrsql{1}{3}{F}{\gamma r \gamma
}{(ij),k,(lm)}\\\stacklrsql{1}{3}{F}{\gamma\gamma\gamma}{(ij),(kl),(mn)}
\end{array}  \right)
\end{array}\right) \,.
\renewcommand{\arraystretch}{1}
\end{equation}

\subsection{The Large-$\bm{D}$ Point Group Symmetry, $\bm{S}_{\bm{N}}$\,, and the Order-By-Order Invariance of the Dimensional Expansion of the Hamiltonian}
\label{subsec:PGS}
The large-$D$ configuration specified by ${\bar{\bm{y}}}_\infty$ has
the highest degree of symmetry where all particles are equidistant
from the center of the trap and equiangular from each other (a
configuration that's only possible in higher dimensions). The point
group is isomorphic to $S_N$ which in effect interchanges the
particles in the large dimension structure. This together with the
fact that the full $D$-dimensional Hamiltonian of
Eqs.~(\ref{eq:scale1})-(\ref{eq:kappa_exp}) is invariant under
particle exchange means the dimensional expansion of
Eq.~(\ref{eq:dpt_exp}) is order-by-order invariant under this $S_N$
point group, i.e.\ the $\bar{H}_j$ are each invariant under the
$S_N$ point group. This greatly restricts the $\bm{F}$ and $\bm{G}$
tensors of Eqs.~(\ref{eq:harm_H}) and (\ref{eq:one_H}). (In three dimensions a corresponding $N$-particle structure
would have a point group of lower symmetry, i.e.\ one not isomorphic to $S_N$ despite the fact that all
of the particles are identical.)

\subsection{The Reducibility of the $\bm{F}$ and $\bm{G}$ Tensors under $\bm{S}_{\bm{N}}$}
\label{subsec:RFGTS}
The maximally symmetric point group $S_N$\,, together with the invariance of the full Hamiltonian
under particle interchange, requires that the $F$
and $G$ tensors be invariant under the interchange of particle
labels (the $S_N$ group). In fact the various blocks of the $F$ and $G$
tensors are themselves invariant under particle interchange induced by the point group.
For example, $\stacklrsql{1}{3}{F}{rr\gamma}{i,j,(kl)}$ is never transformed
into $\stacklrsql{1}{3}{F}{\gamma\gamma\gamma}{(ij),(kl)(mn)}$\,, i.e.\ the $r$
and $\gamma$ labels are preserved since the $S_N$ group does not transform
an $\bar{r}'_i$ coordinate into a $\overline{\gamma}'_{(ij)}$ coordinate.

The various $r-\gamma$ blocks of the $F$ and $G$ tensors may be decomposed into invariant, and, this time irreducible, blocks. Thus for example, $\stacklr{0}{2}{Q}{}{}$ may be decomposed into the blocks
\begin{equation}
\begin{array}{ll}
\stacklr{0}{2}{Q}{rr}{i,i} &
\hspace{2ex} \forall \hspace{2ex} i \\
\stacklr{0}{2}{Q}{rr}{i,j} & 
\hspace{2ex} \forall \hspace{2ex} i \neq j  \\
\stacklr{0}{2}{Q}{r\gamma}{i,(ik)} & \hspace{2ex} \forall
\hspace{2ex} i < k\,,  \\
\stacklr{0}{2}{Q}{r\gamma}{i,(jk)} &  \hspace{2ex} \forall \hspace{2ex}
i \neq j < k \\
\stacklr{0}{2}{Q}{\gamma\gamma}{(ij),(ij)} & \hspace{2ex}
\forall \hspace{2ex} i < j  \\
\stacklr{0}{2}{Q}{\gamma\gamma}{(ij),(ik)} & \hspace{2ex}
\forall \hspace{2ex} i < j \neq k > i \\
\stacklr{0}{2}{Q}{\gamma\gamma}{(ij),(kl)} & \hspace{2ex} \forall
\hspace{2ex} \begin{array}[t]{l} i \neq j \neq k \neq l\,, \\ i <
j\,, \hfill k < l \end{array}
\end{array}
\end{equation}
all of which remain disjoint from one another under the $S_N$ group.
Significantly, invariance under particle interchange requires that tensor elements related by a label permutation induced 
by the point group must be equal. This requirement partitions the set of tensor elements for each block into disjoint
subsets of identical elements. Consider the elements of the $\stackleft{0}{2}Q^{rr}$ block. The element
$\stacklr{0}{2}{Q}{rr}{1,1}$ belongs to a set of $N$ elements (of the form $\stacklr{0}{2}{Q}{rr}{i,i}$) which are
related by a permutation induced by the point group, and therefore must have
equal values. Likewise, the element $\stacklr{0}{2}{Q}{rr}{1,2}$ belongs to a set of $N-1$ elements related by a
permutation induced by the point group and sharing a common value. Proceeding in this fashion, we
observe that the blocks of the harmonic-order tensors are partitioned into the following set of identical
elements which remain disjoint under particle interchange:
\begin{eqnarray}\label{eq:Q0partition}
\stacklr{0}{2}{Q}{rr}{i,i} & = & \stacklr{0}{2}{Q}{rr}{k,k}
\hspace{2ex} \forall \hspace{2ex} i \mbox{ and } k \\
\stacklr{0}{2}{Q}{rr}{i,j} & = & \stacklr{0}{2}{Q}{rr}{k,l}
\hspace{2ex} \forall \hspace{2ex} i \neq j \mbox{ and } k \neq l \\
\stacklr{0}{2}{Q}{r\gamma}{i,(ik)} & = &
\stacklr{0}{2}{Q}{r\gamma}{l,(lm)} =
\stacklr{0}{2}{Q}{r\gamma}{n,(pn)} \hspace{2ex} \forall
\hspace{2ex} i < k\,, \hspace{2ex} l < m\,, \mbox{ and } n
> p \\
\stacklr{0}{2}{Q}{r\gamma}{i,(jk)} & = &
\stacklr{0}{2}{Q}{r\gamma}{l,(mn)} \hspace{2ex} \forall \hspace{2ex}
i \neq j < k \mbox{ and } l
\neq m < n \\
\stacklr{0}{2}{Q}{\gamma\gamma}{(ij),(ij)} & = &
\stacklr{0}{2}{Q}{\gamma\gamma}{(kl),(kl)} \hspace{2ex}
\forall \hspace{2ex} i < j \mbox{ and } k < l \\
\stacklr{0}{2}{Q}{\gamma\gamma}{(ij),(ik)} & = &
\stacklr{0}{2}{Q}{\gamma\gamma}{(lm),(ln)} \hspace{2ex}
\forall \hspace{2ex} i < j \neq k > i  \mbox{ and } l < m \neq n >l \\
\stacklr{0}{2}{Q}{\gamma\gamma}{(ij),(kl)} & = &
\stacklr{0}{2}{Q}{\gamma\gamma}{(mn),(pq)} \hspace{2ex} \forall
\hspace{2ex} \begin{array}[t]{l} i \neq j \neq k \neq l \\ i <
j\,, \hfill k < l \end{array} \mbox{ and }
\begin{array}[t]{l} m \neq n \neq p \neq q \\ m < n\,, \hfill p < q \end{array}
\label{eq:Q0partitiongg}
\end{eqnarray}
%
The harmonic-order block matrices contain the sets of elements in Eqs.~(\ref{eq:Q0partition})-(\ref{eq:Q0partitiongg})
arranged in an intricate pattern. In
Ref.~\onlinecite{FGpaper} it was shown that this arrangement could be expressed in terms of binary matrices formed from matrices
commonly used in graph theory. In this paper, we introduce a generalized structural decomposition for higher orders by
introducing binary tensors, derived using graphs, i.e.\ an extension of binary matrices.

\subsection{Introducing Graphs}
\begin{definition}
A \emph{graph} $\mathcal{G} =(V,E)$ is a set of vertices $V$ and
edges $E$. Each edge has one or two associated vertices, which are
called its \emph{endpoints}.
\end{definition}

For example, \graphggb is a graph $\mathcal{G}$ with three vertices (or ``dots'') and two
edges (or lines). We allow our graphs to include loops and multiple edges\cite{multigraph}.
A graph contains information regarding the connectivity of edges and vertices only: the
orientation of edges and vertices is insignificant. Two graphs with the same number
of vertices and edges that are connected the same way are called \emph{isomorphic}. 

We introduce a mapping which associates each tensor element with a graph as
follows:
\begin{enumerate}
\item draw a labeled vertex ($\labeledgraphv{i}$) for each distinct index in the set of indices of the element
\item draw an edge ($\labeledgraphgamma{i}{j}$) for each double index $(ij)$
\item draw a ``loop'' edge ($\labeledgraphr{i}$) for each distinct single index $i$
\end{enumerate} 
For example, the graph corresponding to the tensor element $\stacklr{0}{2}{Q}{r\gamma}{i,(ij)}$ under this mapping is $\labeledgraphgra{i}{j}$. 

This mapping will be used to represent both the \emph{value} of sets of identical tensor elements and their \emph{tensor structure}. 

The partition of each set of tensor blocks is composed of a certain number of sets of identical elements. Under this mapping, the image of each set of identical elements is a set of isomorphic graphs. We represent each set of isomorphic vertex-labeled graphs (and therefore the corresponding set of identical tensor elements) by an unlabeled graph. 

We denote the set of unlabeled graphs for each block as $\mathbb{G}_{X_1X_2\ldots X_R}$, where $R$ is the rank of the tensor block (and therefore the number of edges in each graph in the set).

\begin{eqnarray}\label{eq:GXX}
\mathbb{G}_{rr}&=&\{\graphrra,\graphrrb\}
\nonumber\\
\mathbb{G}_{\gamma r}&=&\{\graphgra,\graphgrb\}
\\
\mathbb{G}_{\gamma \gamma}
&=&\{\graphgga,\graphggb,\graphggc\}
\nonumber
\end{eqnarray}
\begin{eqnarray}\label{eq:GXXX}
\mathbb{G}_{r}&=&\{\graphr\}
\nonumber\\
\mathbb{G}_{\gamma}&=&\{\graphgamma\}
\nonumber\\
\mathbb{G}_{rrr}&=&\{\graphrrra,\graphrrrb,\graphrrrc\}
\\
\mathbb{G}_{\gamma rr}
&=&\{\graphgrra,\graphgrrb,\graphgrrc,\graphgrrd,\graphgrre\}
\nonumber\\
\mathbb{G}_{\gamma \gamma r}
&=&\{\graphggra,\graphggrb,\graphggrc,\graphggrd,\graphggre,\graphggrf,\graphggrg\}
\nonumber\\
\mathbb{G}_{\gamma \gamma \gamma}
&=&\{\graphggga,\graphgggb,\graphgggc,\graphgggd,\graphggge,\graphgggf,\graphgggg,\graphgggh\}
\nonumber
\end{eqnarray}
These graphs provide a convenient notation to label the sets of tensor elements. We represent the 
\emph{value} of a set of tensor elements by the corresponding unlabeled graph for that set.
For example, the value of the tensor elements $\stacklr{0}{2}{Q}{r\gamma}{i,(ij)}$ is represented
as $\stackleft{0}{2}Q(\graphgra)$. We list the graphs and corresponding elements in
Tables~\ref{tab:ranktwo}, \ref{tab:rankone} and \ref{tab:rankthree}.
Table~\ref{tab:ranktwo} lists the graphs for matrices at harmonic order, along with
the tensor elements represented. 

At the first anharmonic order we have the graphs and the corresponding rank-one and rank-three
Q-tensors in Tables~\ref{tab:rankone} and \ref{tab:rankthree} respectively.

In the next subsection, we represent the \emph{tensor structure} of each set of identical elements corresponding to a graph ${\mathcal G}$ with the binary tensor $B({\mathcal G})$.
\subsection{Decomposition of $Q$ in the Basis of Binary Invariants}
Each tensor block is composed of disjoint sets of identical elements, each labeled by a graph. Each block has a finite number of such sets (and corresponding graphs). We exploit these symmetry properties by introducing a binary tensor for each block and for each graph that embodies the structural arrangement of the corresponding elements. 
\begin{definition}
A \emph{binary invariant} $[B^{block} ({\mathcal G})]_{\nu_1, \nu_2, \ldots}$ corresponding to
a graph ${\mathcal G}$ is a tensor in which elements related to ${\mathcal G}$ by the above mapping are unity and all other elements are zero.
\end{definition}
A binary invariant $B^{block}({\mathcal G})$ of a tensor $\stacklr{order}{R}{Q}{}{}$ has the following properties:
\begin{itemize}
\item The rank of the tensor $B^{block}({\mathcal G})$ equals the number of edges of the graph ${\mathcal G}$.
\item $B^{block}({\mathcal G})$ and $B^{block}({\mathcal G}^\prime)$ are linearly independent, since binary invariants of different graphs share no common elements.
\item The set of binary invariants for the graphs of a block span the unit matrix (that is the sum of a set of binary invariants yields the unit matrix).
\item $B^{block}({\mathcal G})$ is invariant under particle interchange.
\end{itemize}
We introduce a method for deriving expressions for these binary invariants and list them through first-anharmonic order in Ref.~\onlinecite{ongraphs}.

Therefore, we may resolve the $Q$ tensors at any order as a finite linear combination of binary invariants:

\begin{equation}\label{eq:graphresolution}
\stacklr{O}{R}{Q}{block}{\nu_1,\nu_2,\ldots,\nu_R} = \sum_{\mathcal{G}\in
\mathbb{G}_{block}}Q^{block}(\mathcal{G}) \, \left[B^{block}(\mathcal{G})\right]_{\nu_1,\nu_2,\ldots,\nu_R}
\end{equation}
where $\mathbb{G}_{block}$ represents the set of graphs present in the order-$O$, rank-$R$ tensor block
$\stacklr{O}{R}{Q}{block}{}$, and the binary invariant $B^{block}(\mathcal{G})$ has the same dimensions as the original $Q$ tensor block. The scalar quantity $Q^{block}(\mathcal{G})$ is the expansion
coefficient. 

The resolution of symmetric tensor blocks in the basis of binary invariants in Eq.~(\ref{eq:graphresolution}) represents
a generalization of a technique used at harmonic order in Ref.~\onlinecite{FGpaper} to arbitrary order.
This equation also separates the specific interaction dynamics present in $Q^{block}(\mathcal{G})$ from
the point group symmetry embodied in $B^{block}(\mathcal{G})$\,.

\section{Lowest Order: Normal Modes and the Harmonic Wave Function}
\label{sec:NMHWF}
\label{subsec:LowestOrder}
According to Sections~\ref{subsubsec:LDL} and \ref{para:wavef}, the
Jacobian-weighted wave function of spherically confined quantum systems in large dimensions becomes harmonic, with
the system oscillating about the $D \rightarrow \infty$
configuration where every particle is equidistant and equiangular
from every other particle. Notwithstanding the relatively simple
form of the large-dimension, zeroth-order wave function, it includes
beyond-mean-field effects.

>From Eq.~(\ref{eq:harm_H}) one sees that the Schr\"{o}dinger
equation for the zeroth-order Jacobian-weighted wave function has
the form
\begin{equation}\label{Gham}
\left( -\frac{1}{2}\stacklrl{0}{2}{G}{}{\nu_1,\nu_2}
\partial_{{\bar{y}^\prime}_{\nu_1}}
\partial_{{\bar{y}^\prime}_{\nu_2 }} +
\frac{1}{2}\stacklrl{0}{2}{F}{}{\nu_1,\nu_2}
\bar{y}^\prime_{\nu_1 } \bar{y}^\prime_{\nu_2 } +
\stacklrl{0}{0}{F}{}{} \right) \, \Phi_0({\mathbf{\bar{y}'}}) =
\overline{E}_0 \, \Phi_0({\mathbf{\bar{y}'}})
\end{equation}
where ${\bar{\mathbf{y}}'}$ is the dimensionally-scaled displacement
coordinate vector. The matrices (i.e.\ rank-two tensors)
$\stacklru{0}{2}{\mathbf G}{}{}$ and $\stacklru{0}{2}{\mathbf
F}{}{}$ in Eq.~(\ref{Gham}) are both symmetric constant matrices,
while $\stacklru{0}{0}{F}{}{}$ is a scalar quantity.

\subsection{The ${\bf F} \, {\bf G}$ Method}
The Wilson FG method\cite{FGpaper,dcw}
is used to solve equation~(\ref{Gham}) for the normal mode
coordinates and frequencies and allows us to exploit the point group symmetry of the problem.
The $b^{\rm th}$ normal mode coordinate, $q'_b$\,, may be written:
\begin{equation} \label{eq:qyt}
q'_b = V_{b,\nu} \, y'_{\nu} \,.
\end{equation}
The coefficient matrix $\bm{V}$ satisfies the eigenvalue equation
\begin{equation} \label{eq:FGit}
\left[ \stacklru{0}{2}{\mathbf F}{}{} \, \stacklru{0}{2}{\mathbf
G}{}{} \right]_{\nu_1 \nu_2} [V^T]_{\nu_2, b} = \lambda(b) \,
[V^T]_{\nu_1, b}
\end{equation}
and orthonormal condition
\begin{equation} \label{eq:normit}
 V_{a\nu_1} \, V_{b\nu_2} \stacklrl{0}{2}{\mathbf G}{}{}_{\nu_1 \nu_2} = \delta_{a b} \,.
\end{equation}
Eigenvalues, $\lambda(b)$\,, are related to the frequencies,
$\bar{\omega}(b)$\,, by
\begin{equation}\label{eq:omega_b}
\lambda(b)=\bar{\omega}(b)^2.
\end{equation}
In principle, equation~(\ref{eq:FGit}) is still a formidable
equation to solve unless $N$ is quite small since there are $P =
N(N+1)/2$ normal coordinates and up to $P$ distinct frequencies.

\subsection{Group Theory: The $S_N$ Symmetry at Harmonic Order}
As defined in Eq.~(\ref{eq:taylor1}),
the large-dimension structure is completely symmetric, resulting in an $S_N$
point group. Also, as we have noted above, the full $S$-wave Hamiltonian is invariant
under particle interchange, an $S_N$ symmetry under
which the system is invariant.  These two facts mean that Eq.~(\ref{Gham}) is also invariant under the
group $S_N$. This allows us to use the group theory associated with
$S_N$ symmetry which brings about a remarkable reduction from $P$
possible distinct frequencies to five actual distinct frequencies.
The $S_N$ symmetry also greatly simplifies the determination of the
normal coordinates. The vehicle for this is the use of symmetry
coordinates (see below).\cite{dcw}

The fact that $\stacklru{0}{2}{\mathbf F}{}{}$,
$\stacklru{0}{2}{\mathbf G}{}{}$  are invariant matrices under
$S_N$, implies that the $\stacklru{0}{2}{\mathbf F}{}{} \,
\stacklru{0}{2}{\mathbf G}{}{}$ matrix of Eq.~(\ref{eq:FGit}) is
also invariant under $S_N$\,. This means that the
$\stacklru{0}{2}{\mathbf F}{}{} \, \stacklru{0}{2}{\mathbf G}{}{}$
matrix may be expanded in terms of the second rank binary
invariants. The $\stacklru{0}{2}{\mathbf F}{}{}$,
$\stacklru{0}{2}{\mathbf G}{}{}$, and $\stacklru{0}{2}{\mathbf
F}{}{} \, \stacklru{0}{2}{\mathbf G}{}{}$ matrices, which we have
generically denote by $\stacklru{0}{2}{\mathbf Q}{}{}$, are $P
\times P$ matrices which may be written:
\begin{equation}\label{eq:Q}
\stacklru{0}{2}{\mathbf Q}{}{}=\left(\begin{array}{cc}
\stacklru{0}{2}{\mathbf Q}{}{rr} &
\stacklru{0}{2}{\mathbf Q}{}{r \gamma}
\\[0.5ex] \stacklru{0}{2}{\mathbf Q}{}{\gamma r}
& \stacklru{0}{2}{\mathbf Q}{}{\gamma \gamma}
\end{array}\right) = \left(\begin{array}{c|c} \stacklrl{0}{2}{Q}{rr}{i,j}
 & \stacklrl{0}{2}{Q}{r \gamma}{i,(jk)}

 \\[1ex] \hline \\[-1.5ex] \stacklrl{0}{2}{Q}{\gamma r}{(ij),k} & \stacklrl{0}{2}{Q}{\gamma \gamma}{(ij),(kl)} \end{array} \right)
\end{equation}
The upper left block of Eq.~(\ref{eq:Q}) is an $N \times N$ matrix
with elements denoted by indices $i,j$\,. Defining the number of
$\overline{\gamma}'_{ij}$ coordinates to be $M \equiv N(N-1)/2$\,,
the upper right block is an $N \times M$ matrix with elements
denoted by indices $i,(jk)$\,. The lower left block is an $M \times
N$ matrix with elements denoted by indices $(ij),k$\,. The lower
right block is an $M \times M$ matrix with elements denoted by
indices $(ij)\,,(kl)$\,. Using Eq.~(\ref{eq:graphresolution}) we can
write
\begin{equation}\label{GFsub}
\stacklru{0}{2}{\mathbf F}{}{} \, \stacklru{0}{2}{\mathbf G}{}{} =
\left(
\begin{array}{l@{\hspace{3ex}}l}
a \, \bm{B}(\graphrra) + b \, \bm{B}(\graphrrb) & c \,
\bm{B}(\graphgra) + d \, \bm{B}(\graphgrb)
\\[1em]
 e \,
[\bm{B}(\graphgra)]^T + f \, [\bm{B}(\graphgrb)]^T & g \,
\bm{B}(\graphgga) + h \, \bm{B}(\graphggb) + \iota \,
\bm{B}(\graphggc)
\end{array}\right),
\end{equation}
where $a$\,, $b$\, $c$\,, $d$\,, $e$\,, $f$\,, $g$\,, $h$\,, and
$\iota$ are given by Eq.~(42) of Ref.~\onlinecite{FGpaper}\,. The $S_N$
point group symmetry means that the $\stacklru{0}{2}{\mathbf F}{}{} \,
\stacklru{0}{2}{\mathbf G}{}{}$ matrix, instead of having a possible
$P \times P = N(N+1)/2 \times N(N+1)/2$ independent elements, in
fact requires that only {\em nine}\ independent quantities have to
be calculated!

The fact that the $\stacklru{0}{2}{\mathbf F}{}{} \,
\stacklru{0}{2}{\mathbf G}{}{}$ matrix of Eq.~(\ref{eq:FGit}) is an
invariant matrix under $S_N$ further implies that the eigenvectors
${\mathbf{V}}$ of the $\stacklru{0}{2}{\mathbf F}{}{} \,
\stacklru{0}{2}{\mathbf G}{}{}$ matrix and the normal modes
transform under irreducible representations of $S_N$. Using the
theory of group characters\cite{paperI} one sees that the
$r_i$ are reducible to one one-dimensional irreducible
representation labeled by  $[N]$\,, and one
$(N-1)$-dimensional irreducible representation labeled by  $[N-1, \hspace{1ex} 1]$\,. Here, the partition,
$[a_1, \hspace{1ex} a_2, \hspace{1ex} a_3, \hspace{1ex} \ldots \hspace{1ex}  , \hspace{1ex} a_m]$
where $a_1+a_2+a_3+ \cdots +a_m = N$\,, denotes the irreducible representation
of $S_N$\,.\cite{partition, hamermesh}\@ The $\gamma_{ij}$
are reducible to one one-dimensional irreducible
representation labeled by  $[N]$\,, one $(N-1)$-dimensional
irreducible representation labeled by  $[N-1, \hspace{1ex} 1]\,$\,, and one
$N(N-3)/2$-dimensional irreducible representation labeled by  $[N-2, \hspace{1ex} 2]$\,. Since the normal
modes transform under irreducible representations of $S_N$ and are
composed of linear combinations of the elements of the internal
coordinate displacement vectors $\bar{\mathbf{r}}'$ and
$\overline{\mbox{\boldmath $\gamma$}}'$\,, there will be two
$1$-dimensional irreducible representations labeled by the
partition $[N]$, two $(N-1)$-dimensional irreducible representations
labeled by the partition $[N-1, \hspace{1ex} 1]$ and one entirely
angular $N(N-3)/2$-dimensional irreducible representation labeled
by the partition $[N-2, \hspace{1ex} 2]$. All of the normal modes
transforming together under the same irreducible representation will
have the same frequency, and so rather than $N(N+1)/2$ distinct
frequencies there are in fact only five distinct frequencies!

\subsection{The Program to Determine the Normal Modes}
We determine the normal coordinates and distinct frequencies in a
two-step process:\cite{paperI}
\newcounter{twostep}
\newcounter{twostepsecb}
\newcounter{twostepsecc}
\begin{list}{\alph{twostep})}{\usecounter{twostep}\setlength{\rightmargin}{\leftmargin}}
\item \setcounter{twostepsecb}{\value{twostep}}
We start with an important observation: both $\bar{\bm{r}}'$ and
$\overline{\bm{\gamma}}'$ transform under {\em orthogonal}
and, as we have seen above reducible representations of the symmetric group, since both $\bar{\bm{r}}^{\prime \, T} \bm{.} \,\bar{\bm{r}}'$ and $\overline{\bm{\gamma}}^{\prime \, T} \bm{.} \overline{\bm{\gamma}}'$ are invariant under any permutation of particles. Thus ${\bar{\bm{y}}'}$ also
transforms under an {\em orthogonal} reducible representation of the $S_N$
group.

Then we determine two sets of linear combinations of the elements of
coordinate vector $\bar{\bm{r}}'$ which transform under particular
orthogonal $[N]$ and $[N-1, \hspace{1ex} 1]$ irreducible
representations of $S_N$\,. These are the symmetry coordinates for
the $r$ block of the problem\cite{dcw}. Using these
two sets of coordinates we then determine two sets of linear
combinations of the elements of coordinate vector
$\overline{\bm{\gamma}}'$ which transform under exactly the same
orthogonal $[N]$ and $[N-1, \hspace{1ex} 1]$ irreducible
representations of $S_N$ as the coordinate sets in the
$r$ block. Then another set of linear
combinations of the elements of coordinate vector
$\overline{\bm{\gamma}}'$\,, which transforms under a particular
orthogonal $[N-2, \hspace{1ex} 2]$ irreducible representation of
$S_N$\,, is derived. These are then the symmetry coordinates
for the $\gamma$ block of the problem\cite{dcw}.
Furthermore, we choose one of the symmetry coordinates to have the
simplest functional form possible under the requirement that it
transforms irreducibly under $S_N$\,. The succeeding symmetry
coordinate is then chosen to have the next simplest functional
form possible under the requirement that it transforms irreducibly
under $S_N$\,, and so on. In this way the complexity of the
functional form of the symmetry coordinates is kept to a
minimum and only builds up slowly as more symmetry coordinates of
a given species are considered. (More on Step~\alph{twostepsecb}) in Section~\ref{subsubsec:SCW}.)
\item  \setcounter{twostepsecc}{\value{twostep}} The
$\stacklru{0}{2}{\mathbf F}{}{} \, \stacklru{0}{2}{\mathbf G}{}{}$
matrix originally expressed in the $\bar{\mathbf{r}}'$,
$\overline{\mbox{\boldmath $\gamma$}}'$ basis is now expressed in
the symmetry coordinate basis resulting in a remarkable
simplification. The $N(N+1)/2 \times N(N+1)/2$ eigenvalue equation
of Eq.~(\ref{eq:FGit}) is reduced to one $2 \times 2$ eigenvalue
equation for the $[N]$ sector, $N-1$ identical $2 \times 2$
eigenvalue equations for the $[N-1, \hspace{1ex} 1]$ sector and
$N(N-3)/2$ identical $1 \times 1$ eigenvalue equations for the
$[N-2, \hspace{1ex} 2]$ sector. For the $[N]$ and $[N-1,
\hspace{1ex} 1]$ sectors, the $2 \times 2$ structure of the
eigenvalue equations allows for mixing of the symmetry coordinates
in the normal coordinates in the $r$ and
$\gamma$ blocks. The $1 \times 1$
structure of the eigenvalue equations in the $[N-2, \hspace{1ex} 2]$
sector reflects the fact that there are no $[N-2, \hspace{1ex} 2]$
symmetry coordinates in the $r$ block to couple
with, i.e. the $[N-2, \hspace{1ex} 2]$ normal modes are entirely
angular. (More on step~\alph{twostepsecc}) in Section~\ref{subsubsec:NormC}.)
\end{list}

\subsection{Transformation to Symmetry Coordinates and the $W$ Matrices.}
\label{subsubsec:SCW}
Symmetry coordinate vectors ${\bm{S}}_{r}^{\alpha}$ and
${\bm{S}}_{\gamma}^{\beta}$ of the $r$
and $\gamma$ blocks respectively, where $\alpha$
and $\beta$ are the partitions labeling the irreducible
representations of $S_N$ under which the symmetry coordinates
transform, are said to belong to the same species when
$\alpha=\beta$. In what follows we will find it useful to define a
full symmetry coordinate vector as follows:
\begin{equation}\label{eq:FSCV}
{\bm{S}} = \left( \begin{array}{l} {\bm{S}}_{r}^{[N]} \\
{\bm{S}}_{\gamma}^{[N]}  \\ \hline {\bm{S}}_{r}^{[N-1, \hspace{1ex} 1]} \\
{\bm{S}}_{\gamma}^{[N-1, \hspace{1ex} 1]} \\
\hline
{\bm{S}}_{\gamma}^{[N-2, \hspace{1ex} 2]} \end{array} \right) = \left( \begin{array}{l} {\bm{S}}^{[N]} \\
{\bm{S}}^{[N-1, \hspace{1ex} 1]} \\
{\bm{S}}^{[N-2, \hspace{1ex} 2]} \end{array} \right)
 = W \, {\bar{\bm{y}}'} \,,
\end{equation}
where
\begin{equation}\label{eq:srep}
\begin{array}{cc@{\mbox{\hspace{2ex}and\hspace{2ex}}}c}
{\bm{S}}^{[N]} = \left( \begin{array}{l} {\bm{S}}_{r}^{[N]} \\
{\bm{S}}_{\gamma}^{[N]}
\end{array} \right) \,, &
{\bm{S}}^{[N-1, \hspace{1ex} 1]} = \left( \begin{array}{l} {\bm{S}}_{r}^{[N-1, \hspace{1ex} 1]} \\
{\bm{S}}_{\gamma}^{[N-1, \hspace{1ex} 1]}
\end{array} \right) &
\end{array}
{\bm{S}}^{[N-2, \hspace{1ex} 2]} =
{\bm{S}}_{\gamma}^{[N-2, \hspace{1ex} 2]} \,,
\end{equation}
and $W$ is the orthogonal matrix
\begin{equation}\label{eq:W}
W =
\left( \begin{array}{cc} W_{r}^{[N]} & \bm{0} \\
\bm{0} & W_{\gamma}^{[N]} \\
\hline
W_{r}^{[N-1, \hspace{1ex} 1]} & \bm{0} \\
\bm{0} & W_{\gamma}^{[N-1, \hspace{1ex} 1]} \\
\hline \bm{0} & W_{\gamma}^{[N-2, \hspace{1ex} 2]}
\end{array} \right) \,.
\end{equation}
In ${\bm{S}}$, symmetry coordinates of the same species are grouped
together. The $W_{\bm{X}'}^\alpha$\,, where $\alpha = [N]$\,, $[N-1,
\hspace{1ex} 1]$ or $[N-2, \hspace{1ex} 2]$\,, and $\bm{X}'$ is
$r$ or $\gamma$ (only
$\gamma$ for the $[N-2, \hspace{1ex} 2]$ sector),
are given by
\begin{equation}
[W^{[N]}_{r}]_i = \frac{1}{\sqrt{N}} \,
[{\bm{1}}_{r}]_i \;\;\; \mbox{and} \;\;\;
[W^{[N]}_{\gamma}]_{(ij)} = \sqrt{\frac{2}{N(N-1)}}
\,\, [{\bm{1}}_{\gamma}]_{(ij)}
\end{equation}
for the $[N]$ species, where
\begin{equation} \label{eq:bf1i}
[{\bm{1}}_{r}]_i= 1 \;\;\;\; \forall \;\; 1 \leq i \leq N
\;\;\; \mbox{and} \;\;\; [{\bm{1}}_{\gamma}]_{(ij)}
= 1  \;\;\;\; \forall \;\; 1 \leq i,j \leq N \,.
\end{equation}
Regarding the $[N-1, \hspace{1ex} 1]$ species
\begin{equation} \label{eq:WNm1r}
[W^{[N-1, \hspace{1ex} 1]}_{r}]_{ik} =
\frac{1}{\sqrt{i(i+1)}} \left( \sum_{m=1}^i \delta_{mk} - i
\delta_{i+1,\, k} \right) = \frac{1}{\sqrt{i(i+1)}} \left(
\Theta_{i-k+1} - i \,\delta_{i+1,\, k} \right) \,\,,
\end{equation}
where $1 \leq i \leq N-1$ and $1 \leq k \leq N$\,,
\begin{equation}
\begin{array}{r@{\hspace{1ex}}l} {\displaystyle \Theta_{i-j+1} =
\sum_{m=1}^i \delta_{mj} } & =  1 \mbox{ when } j-i<1 \\
& = 0 \mbox{ when } j-i \geq 1 \,, \end{array}
\end{equation}
and
\renewcommand{\jot}{0.5em}
\begin{eqnarray}
\lefteqn{[W^{[N-1, \hspace{1ex}
1]}_{\gamma}]_{i,\,(kl)} = \frac{1}{\sqrt{N-2}}
\,\left( [W^{[N-1, \hspace{1ex} 1]}_{r}]_{ik} \,
[{\bm{1}}_{r}]_l + [W^{[N-1, \hspace{1ex}
1]}_{r}]_{il} \,
[{\bm{1}}_{r}]_k \right) } \nonumber \\
& = & \frac{1}{\sqrt{i(i+1)(N-2)}} \, \bigg( \big( \Theta_{i-k+1} \,
[{\bm{1}}_{r}]_l + \Theta_{i-l+1} \,
[{\bm{1}}_{r}]_k \big) - i \big( \delta_{i+1,\, k} \,
[{\bm{1}}_{r}]_l + \delta_{i+1,\, l} \,
[{\bm{1}}_{r}]_k \big) \bigg) \,, \label{eq:WNm1g}
\end{eqnarray}
\renewcommand{\jot}{0em}
where $[W^{[N-1, \hspace{1ex} 1]}_{r}]_{ik}$ and
$[{\bm{1}}_{r}]_l$ are given by Eqs.~(\ref{eq:WNm1r})
and (\ref{eq:bf1i}) respectively, and $1 \leq k < l \leq N$ and
$1\leq i \leq N-1$\,. Finally, for the $[N-2, \hspace{1ex} 2]$ sector
\begin{equation}
[W^{[N-2, \hspace{1ex} 2]}_{\gamma}]_{(ij),\,(mn)}
= \frac{1}{\sqrt{i(i+1)(j-3)(j-2)}} \,
\begin{array}[t]{r@{\hspace{0.5ex}}l} \Bigl( & (\Theta_{i-m+1} - i
\delta_{i+1,\,m})(\Theta_{j-n} -(j-3)\delta_{jn}) + \\ & +
(\Theta_{i-n+1} - i \delta_{i+1,\,n})(\Theta_{j-m}
-(j-3)\delta_{jm}) \Bigr)  \,, \end{array} \label{eq:WNm2gijmn}
\end{equation}
where $1 \leq i \leq j-2$\,, $4 \leq j \leq N$ and $1 \leq m,n \leq
N$\,.

\subsection{Solution of the $W$-Transformed Eigenequations for the Normal Coordinates}
\label{subsubsec:NormC}
The symmetry coordinate transformed $\stacklru{0}{2}{\mathbf F}{}{}_W
\stacklru{0}{2}{\mathbf G}{}{}_W = W\left(\stacklru{0}{2}{\mathbf F}{}{}
\, \stacklru{0}{2}{\mathbf G}{}{}\right) W^T$ matrix remains invariant
under $S_N$ and so may be expanded in terms of the Clebsch-Gordon
coefficients of $S_N$ coupling two irreducible representations
$\alpha$ and $\beta$ together to form a scalar (i.e.\ invariant) $[N]$
representation, i.e.\ these Clebsch-Gordon coefficients are
invariants in the symmetry coordinate basis. According to group
theory\cite{invariantC}, the only non-zero bilinear invariants which
couple two vectors transforming under orthogonal (more generally
unitary) irreducible representations, couple together equivalent
irreducible representations. Furthermore, there is only one non-zero
bilinear invariant which couples two vectors transforming under
equivalent, orthogonal irreducible representations. As the
irreducible representation is orthogonal, this invariant is
proportional to the Kronecker delta (unit matrix), $\bm{I_{\alpha}}$\,.\cite{harmCG}

Thus we can write
%
%
%
\begin{eqnarray} \label{eq:Qw}
\lefteqn{\stacklru{0}{2}{\mathbf F}{}{}_W \stacklru{0}{2}{\mathbf G}{}{}_W  =}
\nonumber\\&&
 \left( \begin{array}{ccc} \stacklr{0}{2}{\sigma}{{FG}}{[N][N]}
\otimes {\bf I_{[N]}} &
\mathbf{0} & \mathbf{0} 
\\
\mathbf{0} & \stacklr{0}{2}{\sigma}{{FG}}{[N-1, \hspace{1ex} 1][N-1, \hspace{1ex} 1]}
\otimes {\bf I_{[N-1, \hspace{1ex} 1]}} & \mathbf{0} 
\\
\mathbf{0} & \mathbf{0} & \stacklr{0}{2}{\sigma}{{FG}}{[N-2, \hspace{1ex} 2][N-2, \hspace{1ex} 2]} \otimes {\bf I_{[N-2, \hspace{1ex} 2]}}
\end{array} \right) \,.
\end{eqnarray}
We then write Eq.~(\ref{eq:FGit}) in the symmetry coordinate basis
of Eqs.~(\ref{eq:FSCV})--(\ref{eq:WNm2gijmn}) where it reduces to
three eigenvalue equations of the form
\begin{equation} \label{eq:sceig}
\sigmatwo{{FG}}{\alpha}{\alpha}{X_2}{X_1} \,
{[{\mathsf{c}}^{\alpha}]_{Y X_1}} =
\lambda(\alpha,Y) \,
{[{\mathsf{c}}^{\alpha}]_{Y X_2}} \,.
\end{equation}
The reduced $\stacklru{0}{2}{\mathbf F}{}{} \stacklru{0}{2}{\mathbf
G}{}{}$ block matrices $\sigmatwo{{FG}}{[N]}{[N]}{X_2}{X_1}$ and $\sigmatwo{{FG}}{[N-1,1]}{[N-1,1]}{X_2}{X_1}$ are $2 \times 2$-dimensional
matrices ($Y = \pm$\, and $X_1\,, X_2 =
r$ or $\gamma$), while
$\sigmatwo{{FG}}{[N-2,2]}{[N-2,2]}{X_2}{X_1}$ is a one-dimensional matrix
($Y = \gamma$\, and $X_1 = X_2
= \gamma$). The elements of the
$\sigmatwo{{FG}}{\alpha}{\alpha}{X}{X}$ matrices are known analytic
functions specific to the individual systems being considered. Thus
there are five solutions to Eq.~(\ref{eq:sceig}) which we denote as
${\bf 0}^\pm = \{\lambda([N], \pm),\;[{\mathsf{c}}^{[N]}]_\pm\}$\,,
${\bf 1}^\pm = \{\lambda([N-1, \hspace{1ex} 1],
\pm),\;[{\mathsf{c}}^{[N-1, \hspace{1ex} 1]}]_\pm\}$ and ${\bf 2} =
\{\lambda([N-2, \hspace{1ex} 2]),\;{\mathsf{c}}^{[N-2, \hspace{1ex}
2]}\}$\,. The two element $[{\mathsf{c}}^{\alpha}]_\pm$ vectors for
the $\alpha = [N]$ and $[N-1, \hspace{1ex} 1]$ sectors determine the
amount of angular-radial mixing between the symmetry coordinates in
a normal coordinate of a particular $\alpha$\,. Thus
\begin{equation} \label{eq:qS}
[q']_b = {[{\mathsf{c}}^{\alpha}]_\pm}_{r} \,
[{\mathbf{S}}_{r}^{\alpha}]_\xi \, + \,
{[{\mathsf{c}}^{\alpha}]_\pm}_{\gamma} \, [{\mathbf{S}}_{\gamma}^{\alpha}]_\xi \,.
\end{equation}
The normal-coordinate label, $b$, has been replaced by the labels
$\alpha$, $\xi$ and $\pm$ on the right hand side of
Eq.~(\ref{eq:qS}).

>From Eqs.~(\ref{eq:normit}) the ${\mathsf{c}}^{\alpha}$ also satisfy
the orthonormal condition
\begin{equation} \label{eq:cnormeq}
{[{\mathsf{c}}^{\alpha}]_{Y_1 X_1}} \,
{[{\mathsf{c}}^{\alpha}]_{Y_2 X_2}} \,
\sigmatwo{{G}}{\alpha}{\alpha}{X_2}{X_1}
= \delta_{Y_1,
Y_2} \,,
\end{equation}
where the diagonal $\sigmatwo{{G}}{\alpha}{\alpha}{X_2}{X_1}$ matrices are the reduced
$\stacklru{0}{2}{\mathbf G}{}{}$ matrices in the symmetry coordinate
basis.

For the $[N-2, \hspace{1ex} 2]$ sector, the symmetry coordinates are
also the normal coordinates up to a normalization constant,
${\mathsf{c}}^{[N-2, \hspace{1ex} 2]}$ (see Eq.~(\ref{eq:cnormeq}) above).

\subsection{The Wave Function.} \label{subsubsec:wavef}
Thus the wave function of Eq.~(\ref{eq:psiintro}) is the product of $P = N(N+1)/2$ harmonic
oscillator wave functions\cite{paperI}
\begin{equation} \label{eq:Phi__0}
\Phi_0({\mathbf{\bar{y}'}}) = \prod_{\mu=
{\bf 0}^\pm, {\bf 1}^\pm, {\bf 2}}
\psi_{\mu}\left(  {\mathbf {q}}^\prime_{\mu} \right) \,, 
\end{equation}
where $\mu$ labels the manifold of normal modes with the same
frequency $\bar{\omega}_\mu$ and
\begin{equation} \label{eq:psi1D}
\psi_{\mu}({\mathbf {q}}^\prime_{\mu}) = \prod_{i=1}^{d_{\mu}}
\phi_{n_{\mu_i}}\left(  \sqrt{\bar{\omega}_\mu} 
\, q^\prime_{\mu_{i}} \right)
\,, 
\end{equation}
$\phi_{n_{\mu_i}} \left( \sqrt{\bar{\omega}_\mu} 
\, q^\prime_{\mu_{i}} \right) $ is a one-dimensional
harmonic-oscillator wave function of frequency $\bar{\omega}_\mu$
and $n_{\mu_i}$ is the oscillator quantum number, $0 \leq n_{\mu_i}
< \infty$, which counts the number of quanta in each normal mode.
The quantity $d_\mu$ is the number of normal modes in the degenerate manifold of frequency
frequency $\bar{\omega}_\mu$\,, and has values $d_{\mu} = 1$\,, $N-1$ or
$N(N-3)/2$ for $\mu = {\bf 0}^\pm$\,, ${\bf 1}^\pm$ or ${\bf 2}$
respectively.

\section{Next-order: The First Anharmonic Wave Function}
\label{sec:FAWF}
As we have seen in Section~\ref{para:e1}, the next-order, beyond-harmonic energy correction is zero. This though, is not the case for the next-order, beyond-harmonic wave function.
\subsection{Next-Order, Beyond-Harmonic Wave Function}
Expanding  the wave function through first-order beyond harmonic according to Eq.~(\ref{eq:Phit1}),
the wave function correction $\Delta_{1}$, satisfies the commutator equation Eq.~(\ref{eq:deltacommute}). As we have noted, the lowest-order wave function $\Phi_{0} (
\bar{y}')$ (see Eqs.~(\ref{eq:Phi__0}) and (\ref{eq:psi1D})) and Hamiltonian $\bar{H}_{0}$ are
separable in the normal modes $q'_{\mu}$. Thus to most easily
calculate the effect that the next-order Hamiltonian $\bar{H}_1$
(see Eq. (\ref{eq:one_H})) has on the wave function (and energy) we need to
express $\bar{H}_1$ in terms of the normal coordinates.

As in the discussion of the harmonic-order solution, we achieve this
in a two step process, first transforming to symmetry coordinates
(an orthogonal transformation) and then to normal coordinates.

\subsection{Transformation of $\bar{H}_1$ to Symmetry Coordinates}
\label{subsec:H12SC}
>From Eqs.~(\ref{eq:FSCV}) and (\ref{eq:W})
\begin{equation}
 \bar{y}'_{\nu_1} = W^T_{\nu_1,\nu_2} \, S_{\nu_2}
\end{equation}
and
\begin{equation}
 \partial_{y'_{\nu_1}} = W^T_{\nu_1, \nu_2} \, \partial_{S_{\nu_2}} \,,
\end{equation}
where we remind the reader that we are using the repeated index summation convention and
$1 \leq \nu_1, \nu_2 \leq P=N(N+1)/2$\,, then from Eq. (\ref{eq:one_H})
\begin{eqnarray}
 \bar{H}_1 &=& - \frac{1}{2} \, [_3^{(1)} G_W]_{\nu_1, \nu_2, \nu_3}
  S_{\nu_1} \, \partial_{S_{\nu_2}} \, \partial_{S_{\nu_3}} - \frac{1}{2} \, [^{(1)}_1
  G_W]_{\nu} \, \partial_{S_\nu} 
\nonumber\\&&
+ \frac{1}{3!} \, [^{(1)}_3 F_W]_{\nu_1, \nu_2, \nu_3}
  \, S_{\nu_1} \, S_{\nu_2} \, S_{\nu_3} + [^{(1)}_1 F_W]_{\nu} \, S_{\nu} \,, \label{eq:H1Symm}
\end{eqnarray}
where
\begin{eqnarray}
  \label{eq:E8} [^{(1)}_1 G_W]_{\nu} &=& W_{\nu,\eta} \, [^{(1)}_1 G]_{\eta} 
  \\
  \label{eq:E9} [^{(1)}_1 F_W]_{\nu} &=&
  W_{\nu \eta} \,  [^{(1)}_1 F]_{\eta}  \\
  \label{eq:E10} [^{(1)}_3 G_W]_{\nu_1,\nu_2,\nu_3} &=&
  W_{\nu_1,\eta_1} \, W_{\nu_2,\eta_2} \, W_{\nu_3,\eta_3}
  \, [^{(1)}_3 G]_{\eta_1,\eta_2,\eta_3}  \\
  \label{eq:E11} [_3^{(1)} F_W]_{\nu_1,\nu_2,\nu_3} &=&
  W_{\nu_1,\eta_1} \,W_{\nu_2,\eta_2} \, W_{\nu_3,\eta_3} \,
  [^{(1)}_3 F]_{\eta_1,\eta_2,\eta_3}  \,,
\end{eqnarray}
and $1 \leq \eta_1, \eta_2, \eta_3 \leq P$\,.
In Sections~\ref{subsec:PGS} and \ref{subsec:RFGTS} we noted the fact that the $\bm{F}$ and $\bm{G}$
tensors are invariant under $S_N$, leading to the expansion
of these tensors in terms of binary invariants $B(\mathcal{G})$
specified by the graph $\mathcal{G}$ (see Eq. (\ref{eq:graphresolution})).
Thus the transformation of the $F$ and $G$ tensors results from the
transformation properties of the binary invariants.

The third-rank binary invariants transform as
\begin{equation} \label{eq:W3beta}
[B_W (\mathcal{G})]_{\nu_1,\nu_2,\nu_3} =
  W_{\nu_1,\eta_1} \,W_{\nu_2,\eta_2} \, W_{\nu_3,\eta_3} \,
  [B(\mathcal{G})]_{\eta_1,\eta_2,\eta_3}
\end{equation}
while the rank-one binary invariants transform as
\begin{equation} \label{eq:W1beta}
 [B_W (\mathcal{G})]_{\nu} = W_{\nu,\eta} \, [B (\mathcal{G})]_\eta
\end{equation}
Now if $T$ is an element of the $S_N$ group such that
\begin{equation}
{\bar{\bm{y}}}'_{T} = T \,{\bar{\bm{y}}}',
\end{equation}
then the symmetry coordinate vector $\bm{S}$ transforms as
\begin{equation}
\bm{S}_T = T_{W}\,\bm{S}
 \end{equation}
where
\begin{equation}
T_W = W\,T\,W^T \,.
\end{equation}
It follows that the symmetry coordinate transformed binary invariant $B_W
(\mathcal{G})$ is also invariant, coupling three symmetry
coordinates, or one symmetry coordinate, together to give a scalar
term under $S_N$, i.e. one that transforms under the $[N]$
representation.
Since the $\bm{S}$ vector is composed of symmetry coordinates
transforming under irreducible representations of $S_N$ (see Eqs.~(\ref{eq:FSCV}) and (\ref{eq:srep})),
$B_W (\mathcal{G})$ is thus composed of Clebsch-Gordon coefficients
of $S_N$ which couple the irreducible representations together to
form a scalar [N] irrep.
Regarding the rank 1 terms, where there is only one symmetry
coordinate involved, this has to be ${\bm{S}}_{r}^{[N]}$ or
${\bm{S}}_{\gamma}^{[N]}$ since the other symmetry
coordinates are not scalar under $S_N$.
Thus we can write in the block form of Eq.~(\ref{eq:FSCV})
\begin{equation} \label{eq:BW0r}
 B_W (\graphr) = \left( \betaoneG{\bm{0}}{\graphr}{r}, 0, 0, 0, 0 \right) \,,
\end{equation}
and
\begin{equation}  \label{eq:BW0g}
 B_W (\graphgamma) = \left( 0, \betaoneG{\bm{0}}{\graphgamma}{\gamma}, 0, 0, 0 \right) \,,
\end{equation}
where from Appendix~\ref{sec:bincoeff}, Table~\ref{table:sigma1}
\renewcommand{\jot}{1em}
\begin{eqnarray}
\betaoneG{\bm{0}}{\graphr}{r} & = & \sqrt{N} \,,
\\
\betaoneG{\bm{0}}{\graphgamma}{\gamma} & = & \sqrt{\frac{N(N-1)}{2}} \,,
\end{eqnarray}
\renewcommand{\jot}{0em}
so that
\begin{equation}
 B_W (\graphr) \, {\bm{S}} = \betaoneG{\bm{0}}{\graphr}{r} \, {\bm{S}}_{r}^{[N]}
\end{equation}
and
\begin{equation}
 B_W (\graphgamma) \, {\bm{S}} = \betaoneG{\bm{0}}{\graphgamma}{\gamma} \, {\bm{S}}_{\gamma}^{[N]} \,.
\end{equation}

Now consider the rank 3 terms. There are only so many ways we
can couple three irreducible representations drawn from
$[N]$\,, $[N-1,1]$\,, and $[N-2,2]$ irreps.\ to form a scalar $[N]$ irrep.\,. From the
Clebsch-Gordon series for $S_N$ by Murnaghan, Gamba and others\cite{MurnaghanGamba}, these are
\begin{equation}
\begin{array}{r@{\hspace{0.5ex}}c@{\hspace{0.5ex}}l}
  {}[N] \otimes [N] \otimes [N] &  = & [N] + \cdots \\
  {}[N] \otimes [N - 1, 1] \otimes [N - 1, 1] &  = & [N] + \cdots \\
  {}[N] \otimes [N - 2, 2]\otimes [N - 2, 2] &  & [N] + \cdots \\
  {}[N - 1, 1] \otimes [N - 1, 1] \otimes [N - 1, 1] &  = & [N] + \cdots \\
  {}[N - 1, 1] \otimes [N - 1, 1] \otimes [N - 2, 2] & = & [N] + \cdots \\
  {}[N - 1, 1] \otimes [N - 2, 2] \otimes [N - 2, 2] &  = & [N] + \cdots
\end{array}
\end{equation}
and two linearly independent couplings
\begin{equation}
 [N - 2, 2] \otimes [N - 2, 2] \otimes [N - 2, 2] = 2\, [N] + \cdots \,.
\end{equation}
In what follows we denote these two different couplings of three $[N - 2, 2]$
irreps.\ together to form an $[N]$ irrep.\ by the roman numerals for the numbers 1 and 2\,,
i.e.\ $\Rmn{1}$ and $\Rmn{2}$ respectively. Thus we can
write a third rank $B_W (\mathcal{G})$ in block form as\\
\renewcommand{\arraystretch}{1.5}
\begin{equation}
  \label{eq:E19} B_W ({\mathcal G}) = \left(\begin{array}{c@{\hspace{3ex}}c@{\hspace{3ex}}c}
    \left(\begin{array}{c}
      B^{\bm{000}}_W ({\mathcal G})\\
      B^{\bm{010}}_W ({\mathcal G})=0\\
       B^{\bm{020}}_W ({\mathcal G})=0
    \end{array}\right) & \left(\begin{array}{c}
       B^{\bm{001}}_W ({\mathcal G})=0\\
      B^{\bm{011}}_W ({\mathcal G})\\
       B^{\bm{021}}_W ({\mathcal G})=0
    \end{array}\right) & \left(\begin{array}{c}
       B^{\bm{002}}_W ({\mathcal G})=0\\
       B^{\bm{012}}_W ({\mathcal G})=0\\
      B^{\bm{022}}_W ({\mathcal G})
    \end{array}\right) \protect\\[8ex]
    \left(\begin{array}{c}
       B^{\bm{100}}_W ({\mathcal G})=0\\
      B^{\bm{110}}_W ({\mathcal G})\\
       B^{\bm{120}}_W ({\mathcal G})=0
    \end{array}\right) & \left(\begin{array}{c}
      B^{\bm{101}}_W ({\mathcal G})\\
      B^{\bm{111}}_W ({\mathcal G})\\
      B^{\bm{121}}_W ({\mathcal G})
    \end{array}\right) & \left(\begin{array}{c}
       B^{\bm{102}}_W ({\mathcal G})=0\\
      B^{\bm{112}}_W ({\mathcal G})\\
      B^{\bm{122}}_W ({\mathcal G})
    \end{array}\right)\\[8ex]
    \left(\begin{array}{c}
       B^{\bm{200}}_W ({\mathcal G})=0\\
       B^{\bm{210}}_W ({\mathcal G})=0\\
      B^{\bm{220}}_W ({\mathcal G})
    \end{array}\right) & \left(\begin{array}{c}
       B^{\bm{201}}_W ({\mathcal G})=0\\
      B^{\bm{211}}_W ({\mathcal G})\\
      B^{\bm{221}}_W ({\mathcal G})
    \end{array}\right) & \left(\begin{array}{c}
      B^{\bm{202}}_W ({\mathcal G})\\
      B^{\bm{212}}_W ({\mathcal G})\\
      B^{\bm{222}}_W ({\mathcal G})
    \end{array}\right)
  \end{array}\right)
  \,,
\end{equation}
\renewcommand{\arraystretch}{1}
\\
where each block $B^{\alpha_1 \alpha_2 \alpha_3}_W ({\mathcal G})$ is related to the appropriate Clebsch-Gordan coefficient by a scalar multiplier $\beta^{\alpha_1 \alpha_2 \alpha_3,\mathcal{R}} ({\mathcal G})$,
\begin{equation}\label{eq:betaC}
B^{\alpha_1 \alpha_2 \alpha_3}_W ({\mathcal G}) = \sum_{\mathcal{R}} \beta^{\alpha_1
  \alpha_2 \alpha_3, \mathcal{R}} ({\mathcal G}) \,C^{\alpha_1 \alpha_2 \alpha_3,\mathcal{R}}
  \,.
\end{equation}
%
%
In Eq.~(\ref{eq:betaC}), $C^{\alpha_1 \alpha_2
\alpha_3,\mathcal{R}}$ is the appropriate Clebsch-Gordon coefficient
of $S_N$ for the irreps.\,. As indicated above there are
two linearly independent Clebsch-Gordon coefficients coupling three
$[N-2, \hspace{1ex} 2]$ irreps.\ together to form a scalar [N]
irrep. Thus 
$\mathcal{R}$ has two values, running from $\Rmn{1}$ to $\Rmn{2}$ when
$\alpha_1 \alpha_2 \alpha_3 = \mathbf{222}$\,.
All other couplings of the $[N]$\,, $[N-1, \hspace{1ex} 1]$ and
$[N-2, \hspace{1ex} 2]$ irreps.\ have only one Clebsch-Gordon
coefficient, and so $\mathcal{R}$ has only one value, $\Rmn{1}$\,.

One further notes that
\begin{equation}
C^{\alpha_1 \alpha_2 \alpha_3,\mathcal{R}} =
\left(\alpha_1\alpha_2\right) C^{\alpha_1 \alpha_2
\alpha_3,\mathcal{R}}
\end{equation}
where $\left(\alpha_1\alpha_2\right)$ is the operator which
interchanges $\alpha_1$ with $\alpha_2$. Likewise for all of the
other permutations of $\alpha_1, \alpha_2,$and $\alpha_3$. Thus we
have
\begin{equation}
\beta^{\alpha_2 \alpha_1 \alpha_3,\mathcal{R}} ({\mathcal G}) =
\beta^{\alpha_1
  \alpha_2 \alpha_3,\mathcal{R}} ({\mathcal G})
\end{equation}
and likewise for the other permutations of $\alpha_1, \alpha_2,$ and
$\alpha_3$, i.e.\ $\beta^{\alpha_1 \alpha_2 \alpha_3,\mathcal{R}}
({\mathcal G})$ is symmetric in $\alpha_1, \alpha_2,$ and
$\alpha_3$.

The necessary Clebsch-Gordon coefficients are derived in Appendix~\ref{sec:CCGC}, while the multipliers,
$\beta^{\alpha_1 \alpha_2 \alpha_3,\mathcal{R}} ({\mathcal G})$\,, of the  Clebsch-Gordon coefficients
in the expansion of the binary invariants in the normal coordinate basis are derived as analytic functions
of $N$ in Appendix~\ref{sec:bincoeff}. These calculated values, together with the values of the first
anharmonic $G$ and $F$ elements for the particular system of interest, result in a first-anharmonic
order Hamiltonian in the symmetry coordinate basis.

Thus pulling everything together from Eqs.~(\ref{eq:graphresolution}), (\ref{eq:H1Symm})-(\ref{eq:W1beta}),
(\ref{eq:BW0r}), (\ref{eq:BW0g}), and (\ref{eq:betaC}), we find
\renewcommand{\jot}{1em}
\begin{eqnarray}
  \lefteqn{[\stackleft{1}{3}G_W]_{\nu_1,\nu_2,\nu_3} =} \nonumber \\ & & \sum_{\mathcal{G} \in \mathbb{G}^{X_1X_2X_3}}
\stackleft{1}{3}G^{X_1X_2X_3}(\mathcal{G}) \, [B_{W}^{\alpha_1\alpha_2\alpha_3}(\mathcal{G})]_{\xi_1,\xi_2,\xi_3}
= \sum_{\mathcal{R}}
\sigmathree{G}{\alpha_1}{\alpha_2}{\alpha_3,\mathcal{R}}{X_1}{X_2}{X_3}
\, C^{\alpha_1\alpha_2\alpha_3,\mathcal{R}}_{\xi_1,\xi_2,\xi_3} \,, \hspace{3ex} \label{eq:GW13C}
\end{eqnarray}
\begin{eqnarray}
  \lefteqn{[\stackleft{1}{3}F_W]_{\nu_1,\nu_2,\nu_3} =}  \nonumber \\ && \sum_{\mathcal{G} \in \mathbb{G}^{X_1X_2X_3}}
\stackleft{1}{3}F^{X_1X_2X_3}(\mathcal{G}) \, [B_{W}^{\alpha_1\alpha_2\alpha_3}(\mathcal{G})]_{\xi_1,\xi_2,\xi_3}
= \sum_{\mathcal{R}}
\sigmathree{F}{\alpha_1}{\alpha_2}{\alpha_3,\mathcal{R}}{X_1}{X_2}{X_3}
\, C^{\alpha_1\alpha_2\alpha_3,\mathcal{R}}_{\xi_1,\xi_2,\xi_3} \,, \hspace{3ex} \label{eq:FW13C}
\end{eqnarray}
\renewcommand{\jot}{0em}
\renewcommand{\jot}{1em}
\begin{eqnarray}
\label{eq:sigma11G}
[^{(1)}_1 G_W]_{\nu} = \sum_{\mathcal{G}\in\mathbb{G}_{X}} \stacklr{1}{1}{G}{X}{}\!\!\left(\mathcal{G}\right)
\, [B_W^{\alpha}(\mathcal{G})]_\xi
& = & \sigmaone{G}{\alpha}{X} \, C^\alpha_\xi
 \,,
\\
{} [^{(1)}_1 F_W]_{\nu} = 
\sum_{\mathcal{G}\in\mathbb{G}_{X}}\stacklr{1}{1}{F}{X}{}\!\!\left(\mathcal{G}\right)
\, [B_W^{\alpha}(\mathcal{G})]_\xi 
& = & \sigmaone{F}{\alpha}{X} \, C^\alpha_\xi \,, \label{eq:sigma11F}
\end{eqnarray}
\renewcommand{\jot}{0em}
where
\renewcommand{\jot}{1em}
\begin{eqnarray}
\label{eq:sigma13G}
\sigmathree{G}{\alpha_1}{\alpha_2}{\alpha_3,\mathcal{R}}{X_1}{X_2}{X_3} &=&\sum_{\mathcal{G}\in\mathbb{G}_{X_1X_2X_3}}\stacklr{1}{3}{G}{X_1X_2X_3}{}\left(\mathcal{G}\right)
\,\betathreeG{\alpha_1}{\alpha_2}{\alpha_3,\mathcal{R}}{\mathcal G}{X_1}{X_2}{X_3} \,,
\\
\sigmathree{F}{\alpha_1}{\alpha_2}{\alpha_3,\mathcal{R}}{X_1}{X_2}{X_3} &=&\sum_{\mathcal{G}\in\mathbb{G}_{X_1X_2X_3}}\stacklr{1}{3}{F}{X_1X_2X_3}{}\left(\mathcal{G}\right)
\,\betathreeG{\alpha_1}{\alpha_2}{\alpha_3,\mathcal{R}}{\mathcal G}{X_1}{X_2}{X_3} \,. \label{eq:sigma13F}
\\
\sigmaone{G}{\alpha}{X} &=&\sum_{\mathcal{G}\in\mathbb{G}_{X}}\stacklr{1}{1}{G}{X}{}\left(\mathcal{G}\right)
\,\betaoneG{\alpha}{\mathcal G}{X}
\nonumber\\
\sigmaone{F}{\alpha}{X} &=&\sum_{\mathcal{G}\in\mathbb{G}_{X}}\stacklr{1}{1}{F}{X}{}\left(\mathcal{G}\right)
\,\betaoneG{\alpha}{\mathcal G}{X}
\end{eqnarray}
\renewcommand{\jot}{0em}
We also have
\begin{equation}
C^\alpha_1 = 1  \mbox{\hspace{3ex}when\hspace{1ex}}
\alpha = \bm{0}\,, \mbox{\hspace{1ex}and\hspace{1ex}}
C^\alpha_\xi = 0  \mbox{\hspace{3ex}when\hspace{1ex}}
\alpha = \bm{1}\,, \mbox{\hspace{1ex}or\hspace{1ex}} \bm{2} \,,
\end{equation}
since, from Eqs.~(\ref{eq:BW0r}) and (\ref{eq:BW0g}), $B_W^{\alpha}(\mathcal{G}) = 0$ when $\alpha = \bm{1}$ or $\bm{2}$\,.
In Eqs.~(\ref{eq:GW13C})-(\ref{eq:sigma13F}) above, repeated $\alpha_1$\,, $\alpha_2$\,, $\alpha_3$\,,
$X_1$\,, $X_2$\,, $X_3$\,, and $X$ labels are {\em not} summed over. We remind the reader that
$X_1$\,, $X_2$\,, $X_3$\,, and $X$ label $r$ or $\gamma$\,; $\alpha_1$\,, $\alpha_2$\,, $\alpha_3$\,, and $\alpha$
label the irreducible representation of $S_N$ ($[N] = \bm{0}$\,, $[N-1, \hspace{1ex} 1] = \bm{1}$\,, or
$[N-2, \hspace{1ex} 2] = \bm{2}$). The labels $\nu_1$\,, $\nu_2$\,, $\nu_3$\,, and $\nu$ cover both
$\alpha_1$\,, $\alpha_2$\,, $\alpha_3$\,, $\alpha$\,, and $X_1$\,, $X_2$\,, $X_3$\,, $X$ as well as the irrep.\
element labels $i$\,, $j$\,, $(kl)$\,, etc.\,. A summary of our use of different label sets can be found in Ref.~\onlinecite{ongraphs}.

\subsection{Transformation To Normal Coordinates}
\label{sec:transnorm}
The transformation to normal coordinates is
affected by a transformation which only mixes the $r$ and
$\gamma$ blocks i.e.
\begin{equation}
q^{\alpha} = \bm{\mathcal{C}}^{\alpha}\,\bm{S}^{\alpha}
\end{equation}
where $\bm{S}^{\alpha}$ is given by Eq. (\ref{eq:srep}).The matrices $\bm{\mathcal{C}^{0}}$ and $\bm{\mathcal{C}^1}$ are $2\times 2$ matrices, while $\bm{\mathcal{C}^{2}}$ is a scalar quantity. \\

For the $\bm{0}$ and $\bm{1}$ sectors
\begin{equation}
\bm{\mathcal{C}}^{\alpha} =
\renewcommand{\arraystretch}{1.5}\left(\begin{array}{l@{\hspace{4ex}}l}
C^{\alpha}_{+}\cos\theta^{\alpha}_{+} & C^{\alpha}_{+}\sin\theta^{\alpha}_{+}\\
C^{\alpha}_{-}\cos\theta^{\alpha}_{-} &
C^{\alpha}_{-}\sin\theta^{\alpha}_{-}
\end{array}\right)
\renewcommand{\arraystretch}{1}
\end{equation}

Thus
\begin{equation}
\bm{S}^{\alpha} = (\bm{\mathcal{C}}^{\alpha})^{-1}\bm{q}^{\alpha}
\end{equation}

and
\begin{equation}
\frac{\partial}{\partial{\bm{S}^{\alpha}}} =
\frac{\partial}{\partial{\bm{q}^{\alpha}}}\bm{\mathcal{C}}^{\alpha}
\end{equation}

Defining 
\begin{eqnarray}
\label{eq:tau11}
\tauone{G}{\alpha}{Y} &=&
{\cal C}_{Y,X}\sigmaone{G}{\alpha}{X}
\nonumber\\
\tauone{F}{\alpha}{Y} &=&
[{\cal C}^{-1}]^T_{Y,X}\sigmaone{F}{\alpha}{X}
\\
\label{eq:tau13}
\tauthree{G}{\alpha_1}{\alpha_2}{\alpha_3,\mathcal{R}}{Y_1}{Y_2}{Y_3} &=&
[{\cal C}^{-1}]^T_{Y_1,X_1}{\cal C}_{Y_2,X_2}{\cal C}_{Y_3,X_3}\sigmathree{G}{\alpha_1}{\alpha_2}{\alpha_3,\mathcal{R}}{X_1}{X_2}{X_3}
\nonumber\\
\tauthree{F}{\alpha_1}{\alpha_2}{\alpha_3,\mathcal{R}}{Y_1}{Y_2}{Y_3} &=&
[{\cal C}^{-1}]^T_{Y_1,X_1} [{\cal C}^{-1}]^T_{Y_2,X_2} [{\cal C}^{-1}]^T_{Y_3,X_3}\sigmathree{F}{\alpha_1}{\alpha_2}{\alpha_3,\mathcal{R}}{X_1}{X_2}{X_3}
\nonumber\,,
\end{eqnarray}
and defining $G_V$ and $F_V$ to be the $G$ and $F$ tensors in the normal coordinate basis, 
\begin{eqnarray}\label{eq:QV1}
\left[\stacklr{1}{1}{G}{}{V}\right]_{\nu} &=&
\tauone{G}{\alpha}{Y} \,,
\nonumber\\
\left[\stacklr{1}{1}{F}{}{V}\right]_{\nu} &=&
\tauone{F}{\alpha}{Y} \,,
\\
\left[\stacklr{1}{3}{G}{}{V}\right]_{\nu_1,\nu_2,\nu_3} &=&
\sum_\mathcal{R} \tauthree{G}{\alpha_1}{\alpha_2}{\alpha_3,\mathcal{R}}{Y_1}{Y_2}{Y_3}
\,\left[C^{\alpha_1 \alpha_2 \alpha_3,\mathcal{R}}\right]_{\xi_1,\xi_2,\xi_3} \,,
\nonumber\\
\left[\stacklr{1}{3}{F}{}{V}\right]_{\nu_1,\nu_2,\nu_3} &=&
\sum_\mathcal{R} \tauthree{F}{\alpha_1}{\alpha_2}{\alpha_3,\mathcal{R}}{Y_1}{Y_2}{Y_3}
\,\left[C^{\alpha_1 \alpha_2 \alpha_3,\mathcal{R}}\right]_{\xi_1,\xi_2,\xi_3}
\nonumber\,,
\end{eqnarray}
where repeated $\alpha_1$\,, $\alpha_2$\,, and $\alpha_3$
labels are {\em not} summed over,
we obtain $\bar{H}_1$ in normal coordinates
\begin{eqnarray}\label{eq:H1V}
\bar{H}_1&=&-\frac{1}{2}[\stackleft{1}{3}G_V]_{\nu_1,\nu_2,\nu_3}\bar{q}^{\prime}_{\nu_1 }\partial_{\bar{q}^{\prime}_{\nu_2 }}\partial_{\bar{q}^{\prime}_{\nu_3}}
 +\frac{1}{3!}[\stackleft{1}{3}F_V]_{\nu_1,\nu_2,\nu_3}\,\bar{q}^{\prime}_{\nu_1 } \bar{q}^{\prime}_{\nu_2 } \bar{q}^{\prime}_{\nu_3 } 
\nonumber\\&&
 -\frac{1}{2}[\stackleft{1}{1}G_V]_\nu \,\partial_{\bar{q}^{\prime}_\nu}+[\stackleft{1}{1}F_V]_\nu \,\bar{q}^{\prime}_\nu 
\,,
\end{eqnarray}

>From Eqs.~(\ref{eq:Phit1}) and (\ref{eq:deltacommute}),

\begin{equation}
\tag{\ref{eq:Phit1}}
\Phi = (1 + \delta^{1 / 2} \Delta_{1}) \Phi_{0} + O (\delta) \,,
\end{equation}
%
%
%
\begin{equation}
\tag{\ref{eq:deltacommute}}
[\Delta_{1}\,, \, \bar{H}_{0}] \, \Phi_{0} = \bar{H}_{1}
\Phi_0 \,,
\end{equation}

and (\ref{eq:H1V}), we can derive the first-order, beyond-harmonic wave
function.

\subsection{An Example: The First-Order, Beyond-Harmonic Wave Function for the $N$-boson Ground State}

Now
\begin{equation}
\frac{\partial}{\partial{\bar{\bm{q}}^\prime}}\phi_{0}(\omega^{\frac{1}{2}}\bm{q}^\prime) = -\bar{\omega} \, \bm{q}^\prime \,
\phi_{0}(\bar{\omega}^{\frac{1}{2}}\bm{q}^\prime)
\end{equation}
and
\begin{equation}
\frac{\partial^{2}}{\partial{\bm{q}^\prime}^{2}}\phi_{0}(\bar{\omega}^{\frac{1}{2}}\,\bm{q}^\prime) =
\bar{\omega}(\bar{\omega} {\bm{q}^\prime}^{2} - 1)\phi_{0}(\bar{\omega}^{\frac{1}{2}}\,\bm{q}^\prime) \,.
\end{equation}

Therefore from Eq.~(\ref{eq:H1V}) we find that the action of $\bar{H}_1$ on the lowest-order, ground-state wave function ${}_g\Phi_0$ becomes equivalent to the action of a 3rd-order polynomial $\left(\bar{H}_1\right)_{\textrm{eff}}$ :
\begin{equation} \label{eq:H1eff}
\bar{H}_1 \,\, {}_g\Phi_0 = \left(\bar{H}_1\right)_{\textrm{eff}} \,\, {}_g\Phi_{0}
\end{equation}
where
\begin{eqnarray}
\label{eq:H1polynomial}
 \left(\bar{H}_1\right)_{\textrm{eff}} &=& \left(-\frac{1}{2} \left[\stacklr{1}{3}{G}{}{V}\right]_{\nu_1, \nu_2, \nu_3} 
   \bar{\omega}_{\nu_2}  \bar{\omega}_{\nu_3} + \frac{1}{3!}
   \left[\stacklr{1}{3}{F}{}{V}\right]_{\nu_1, \nu_2, \nu_3}\right) {q'}_{\nu_1}
   {q'}_{\nu_2} {q'}_{\nu_3}
\nonumber\\&& + \left(\frac{1}{2} \left[\stacklr{1}{3}{G}{}{V}\right]_{\nu_1, \nu_2, \nu_2} 
   \bar{\omega}_{\nu_2}+\frac{1}{2} \left[\stacklr{1}{1}{G}{}{V}\right]_{\nu_1}  \bar{\omega}_{\nu_1} + \left[\stacklr{1}{1}{F}{}{V}\right]_{\nu_1}  
    \right){q'}_{\nu_1} . 
\end{eqnarray}
We use the index $\mu$ to refer to both the partition $\alpha$ and the block $Y$:
$\mu\in \{\mathbf{0}+,\mathbf{0}+,\mathbf{1}+,\mathbf{1}-,\mathbf{2}\}$.
This is a convenient index to refer to the $5$ blocks of the normal mode vector
$\bar{q}^{\prime}$. In this notation, the elements of the block $q^{\prime}_\mu$
are denoted $\left[q^{\prime}_\mu\right]_\xi$, with the range of $\xi$ implied by $\mu$.

Using this notation, we define the tensor blocks $\stacklr{1}{3}{\tau}{{H_1}}{\mu_1, \mu_2 ,\mu_3,\mathcal{R}}$ and
$\stacklr{1}{1}{\tau}{{H_1}}{\mu_1}$:
\begin{eqnarray}
  \stacklr{1}{3}{\tau}{{H_1}}{\mu_1, \mu_2 ,\mu_3,\mathcal{R}}
&=&
 -\frac{1}{2}\stacklr{1}{3}{\tau}{G}{\mu_1, \mu_2 ,\mu_3,\mathcal{R}}
  {\omega}_{\mu_2}  {\omega}_{\mu_3} + \frac{1}{3!}
  \stacklr{1}{3}{\tau}{F}{\mu_1, \mu_2 ,\mu_3,\mathcal{R}} \label{eq:tau3H1}
\\
\stacklr{1}{1}{\tau}{{H_1}}{\bm{0\pm}}&=&
\frac{1}{2} \sum_{\mu} d_{\mu}
\stacklr{1}{3}{\tau}{G}{\bm{0\pm},\mu,\mu}
\omega_{\mu}+\frac{1}{2} \stackleft{1}{1}\tau_{\bm{0\pm}}^G  \omega_{\bm{0\pm}} + \stackleft{1}{1}\tau_{\bm{0\pm}}^F
 \label{eq:tau1H1}
\end{eqnarray}
where $d_{\mathbf{0\pm}} = 1$\,, $d_{\mathbf{1\pm}} = N-1$, and $d_{\mathbf{2}} = N(N-3)/2$.

>From Eqs.~(\ref{eq:deltacommute}) and (\ref{eq:H1polynomial})

\begin{equation}\label{eq:tauDelta}
\Delta_{1}=\sum_{\renewcommand{\arraystretch}{0.5} \begin{array}{c} \scriptstyle \mu_1,\mu_2, \mu_3 \\ \scriptstyle \xi_1,\xi_2,\xi_3 \\ \scriptstyle \mathcal{R}  \end{array} \renewcommand{\arraystretch}{1} }
\left(
\stacklr{1}{3}{\tau}{\Delta}{\mu_1,\mu_2,\mu_3, {\mathcal{R}}}C^{\mu_1\mu_2\mu_3,{\mathcal{R}}}_{\xi_1,\xi_2,\xi_3}
\right)
[q^\prime_{\mu_1}]_{\xi_1}[q^\prime_{\mu_2}]_{\xi_2}[{q'}_{\mu_3}]_{\xi_3}+ \sum_{Y = \pm} \stacklr{1}{1}{\tau}{\Delta}{\mathbf{0}^Y}{q^\prime}_{\mathbf{0}^Y},
\end{equation}
where
\begin{eqnarray}
\stackleft{1}{3}\tau^\Delta_{\mu_1,\mu_2,\mu_3,\mathcal{R}}
&=&
\frac{-\stackleft{1}{3}\tau^{H_1}_{\mu_1,\mu_2,\mu_3,\mathcal{R}}}{\bar{\omega}_{\mu_1}+\bar{\omega}_{\mu_2}+\bar{\omega}_{\mu_3}}
 \\
 \stackleft{1}{1}{\tau}^{\Delta}_{\mathbf{0}\pm}
 &=&
 \frac{1}{\bar{\omega}_{\mathbf{0}\pm}}
\left(-\stacklr{1}{1}{\tau}{H_1}{\mathbf{0}\pm}
+ \sum_\mu d_\mu
\left(
\stacklr{1}{3}{\tau}{\Delta}{\mathbf{0}\pm\mu\mu}
+\stacklr{1}{3}{\tau}{\Delta}{\mu\mathbf{0}\pm\mu}
+\stacklr{1}{3}{\tau}{\Delta}{\mu\mu\mathbf{0}\pm}
\right)
\right) \label{eq:taud0pm}
\end{eqnarray}

Notice that the next-order wave function is equal to $_g\Phi_{0}$
times a polynomial of odd powers of order three. In Eqs.~(\ref{eq:tau3H1})-(\ref{eq:taud0pm})
repeated indices are {\em not} summed over unless explicitly indicated.

\section{Summary and Conclusions}
\label{sec:SumCon}

The $N$-body problem is ubiquitous in physics as well as chemistry, biology and
related fields. Due to the exponential $N$
scaling of these problems, their solutions
require approximations.
Our work suggests that it may be
advantageous to compartmentalize the $N$ scaling problem away from the rest of
the physics.  We have shown that this may be done
for a system of $N$ identical bosons, using group
theory and graphical methods.  The $N$ scaling is checked due to the structure
imposed at each order of the perturbation series
from the invariance under
the large-dimension point group.  Having identical particles is a necessary, but
not sufficient condition to take full advantage of the group theory of the
symmetric group.  The interparticle distances and angles must be identical,
i.e. a point group achievable only in higher 
dimensions.\cite{FGpaper,loeser}  This
point group in high spatial dimensions is
isomorphic with the symmetric group, $S_N$.

As we have seen above in the body of this paper, higher-order terms
involve a
polynomial in coordinates quantifying the deviation of the system
from this lowest-order configuration. 
The coefficients weighting 
individual monomial terms in these
polynomials may be gathered together to form tensors in the particle 
label space,
where the particle label $i$ runs over $1 \leq i \leq N$\,. Since these tensors are
invariant under the $N!$ operations of the $S_N$ group, they
yield to an exact decomposition in terms of a small basis of binary tensors,
invariant under $S_N$. The basis of binary invariants is finite,
independent of $N$, and has been derived using graphical techniques. Thus the problem and its solution involve a small, $N$-independent, number of coefficients
for this basis. We thus now have a problem which scales as $N^0$\, where $N$ enters
as a parameter.



The result of the analysis outlined in this paper is the wave function 
for a general, interacting system, exact and analytic through first 
order. This first-order correction introduces anhamonicities and couples 
the breathing, single-particle and phonon mode coordinates together, 
which for the ground state takes the form of a cubic polynomial of odd 
powers of the normal mode coordinates multiplying the lowest order 
solution (see Eqs.~(\ref{eq:Phit1}) and (\ref{eq:tauDelta})).

In principle any observable quantity of a confined quantum system
can be derived from this correlated wave function. The
 density profile of a macroscopic confined
quantum system offers an experimentally accessible window into
beyond-mean-field effects in a quantum object. These effects include
a greater spatial extent resulting from hard collisions within the
system as well as fermionization and crystallization in quasi-one
and two dimensional systems\cite{fermcryst}. Results
for the density profile through next-order for large $N$ and/or
large interparticle interaction strength may be derived.\cite{density1harm}
 One notes that the derived Clebsch-Gordon coefficients couple up to three
normal modes together to form a scalar $[N]$ function and this will
allow excited-state wave functions of up to three quanta to be
derived for systems which have completely spatially symmetric wave
functions, i.e. as a spinless bosonic system (e.g.\ a BEC). The fact
that excited states may be derived means that thermal properties at
finite temperatures for cold confined quantum systems can be
calculated in an appropriate thermal sum over states. 

This method generalizes to higher order. Successive terms in the
perturbation series will require a finite number of binary invariants
as a basis
for the tensor blocks at each order.  The number of binary invariants at a
particular order is independent of $N$,  thus checking the $N$ scaling.
However, there is still significant
analytical work
to be done in deriving the Clebsch-Gordon coefficients at each order.  
This work is only possible through the use of the group
theory of the symmetric group which greatly constrains the structure of the
problem and restricts the number of coefficients. Although the work needed will
increase at each order, it remains analytic and will be
a systematic extension of
the work outlined in this paper for the first-order correction to the
wave function. At all orders,  $N$ remains a parameter. Because the
N scaling part of the problem has been compartmentalized away from the
rest of the physics, this work does not need to be repeated for different
$N$ or for a different interaction potential.

Also
the derived Clebsch-Gordon coefficients allow this work for the
correlated next-order wave function for systems with spherical
symmetry to be extended to $m=0$ systems with cylindrical symmetry,
such as most laboratory BECs (the additional normal modes associated
with the $z$ direction transform under the $[N]$ and $[N-1,
\hspace{1ex} 1]$ irreps.\ of the $S_N$ group). Furthermore, the fact that
the $S_N$ point group allows higher orders to be calculated
also paves the way for an investigation of the above mentioned
phenomenon of fermionization and crystallization\cite{fermcryst}, which display
structure in the density profile which is absent from the correlated
lowest-order density profile. Although, the present paper
addresses $L=0$ confined quantum systems, in principle the formalism
has been developed\cite{higherL1,higherL2} which would allow the extension of this work to
higher angular momentum states.

Systems of particles without a confining potential, but which bind under their own interaction may be treated in a similar fashion, in Jacobi coordinates for example.

\bigskip
\bigskip
\bigskip

\section{Acknowledgments}

We gratefully acknowledge continued support from the Army Research Office.  We thank
David Kelle for advice on the use of graphs.

\appendix

\section{Identities}\label{app:identities}
\begin{equation}
\frac{\partial}{\partial \delta^{1/2}} = \bar{y}^\prime_{\nu}
\frac{\partial}{\partial {\bar{y}}_{\nu}} + \frac{d}{d
\delta^{1/2}}
\end{equation}
\begin{equation}
\left(\td{}{\delta^{1/2}}\right)^2 =
 \bar{y}^\prime_{\nu_1} \, \bar{y}^\prime_{\nu_2} \,
\frac{\partial}{\partial {\bar{y}_{\nu_1}}} \,
\frac{\partial}{\partial {\bar{y}_{\nu_2}}} + \, 2 \,
\bar{y}^\prime_\nu \frac{\partial}{\partial {\bar{y}_\nu}}
\frac{d}{d \delta^{1/2}} + \left(\frac{d}{d \delta^{1/2}}\right)^2
\end{equation}
\begin{eqnarray}
\left(\td{}{\delta^{1/2}}\right)^3
& = &
\bar{y}^\prime_{\nu_1} \, \bar{y}^\prime_{\nu_2} \,
\bar{y}^\prime_{\nu_3} \, \frac{\partial}{\partial
\bar{y}_{\nu_1}} \, \frac{\partial}{\partial \bar{y}_{\nu_2}} \,
\frac{\partial}{\partial \bar{y}_{\nu_3}} \\
&& + \, 3\,\bar{y}^\prime_{\nu_1} \, \bar{y}^\prime_{\nu_2} \,
\frac{\partial}{\partial \bar{y}_{\nu_1}} \,
\frac{\partial}{\partial \bar{y}_{\nu_2}} \frac{d}{d \delta^{1/2}}
+ \, 3 \, \bar{y}^\prime_\nu \frac{\partial}{\partial
\bar{y}_\nu} \left(\frac{d}{d \delta^{1/2}}\right)^2 + \left(\frac{d}{d \delta^{1/2}}\right)^3
\end{eqnarray}

\section{Derivation of the Clebsch-Gordon Coefficients}
\label{sec:CCGC}
Deriving the first Clebsch-Gordon coefficient $C^{000,\Rmn{1}}$ is straightforward enough.
It is simply
\begin{equation} C^{000,\Rmn{1}} = 1\,. \end{equation}

Strictly speaking, Clebsch-Gordon coefficients effect a unitary
transformation and satisfy the normalization condition.
\begin{equation}
C^{\alpha_1\alpha_2\alpha_3,\mathcal{R}}_{\xi_1,\xi_2\,\xi_3}C^{\alpha_1\alpha_2\alpha_3,\mathcal{R}}_{\xi_1,\xi_2\,\xi_3}
= 1.
\end{equation}
We do not, however, need the above normalization relationship: we only
need the coefficients to be linearly independent and to span the required space. Therefore, we drop the unitarity requirement and use unnormalized Clebsch-Gordon coefficients.

Now according to the general theory of finite groups (Hamermesh,
Section 5-4, p. 136), there is only one bilinear invariant of an
irreducible representation. If this irrep. is unitary the invariant
is proportional to the Kronecker delta function $\delta_{ij}$\,.
This invariant couples two irreps. together to give a scalar $[N]$ representation. \\

Thus we have
\begin{equation} C^{110,\Rmn{1}}_{ij} = \delta_{ij}\end{equation}
and
\begin{equation}C^{220,\Rmn{1}}_{(ij)(kl)}=\delta_{ik}\delta_{jl}\,,\end{equation}
where $1\le i \le N-1$ and $1 \le j \le N-1$.

More generally we can calculate the required Clebsch-Gordon coefficients by transforming the binary invariants $B({\mathcal G})$ to symmetry coordinates with the required $W$ matrices. 
We perform this and the remaining derivations in this Appendix symbolically using \textsc{Mathematica}\cite{mathematica}. 

Using the proportionality relation
\begin{equation} C^{111,\Rmn{1}}_{ijk} 
\propto [W^{1}_{r}]_{il}[W^{1}_{r}]_{jm}[W^{1}_{r}]_{kn}[B({\mathcal G})]_{lmn}
\,,
\end{equation}
we define the unnormalized coefficient
\begin{eqnarray} 
\lefteqn{C^{111,\Rmn{1}}_{ijk} =\frac{1}{\sqrt{i (i+1) j (j+1) k (k+1)}}\left(-i (i^2-1) \delta_{ij}\delta_{ik}\right.}\nonumber\\
&&\left.+i (i+1) \Theta _{k-i} \delta_{i,j}+k (k+1) \Theta _{j-k} \delta_{i,k}+j (j+1) \Theta _{i-j} \delta_{j,k}\right)\,,
\end{eqnarray}
where  $1\le i \le N-1$, $1\le j \le N-1$, and $1\le k \le N-1$.

Similarly, using the proportionality relation
\begin{equation} 
C^{211,\Rmn{1}}_{(ij),k,l} 
\propto [W^{2}_{\gamma}]_{(ij)m}[W^{1}_{r}]_{kn}[W^{1}_{r}]_{lp}[B({\mathcal G})]_{mnp}
\,,
\end{equation}
we obtain the coefficient
\begin{eqnarray} 
\lefteqn{C^{211,\Rmn{1}}_{(ij),k,l} =
\frac{1}{\sqrt{i(i+1)(j-3)(j-2)k(k+1)l(l+1)}}}
\nonumber\\&&
\times\left(2i(i+1)(\Theta_{-j+l+1}-\Theta_{l-k})\delta_{i,k}+2i(i+1)(\Theta_{-j+k+1}-\Theta_{k-l})\delta_{i,l}
\right.\nonumber\\*&&\left.+i(i+1)(j-2)(j-1)(\delta_{i,l}\delta_{j-1,k}+\delta_{i,k}\delta_{j-1,l})-2l(l+1)\Theta_{i-k}\delta_{k,l}
\right.\nonumber\\&&\left.+2i(i^2-1)\delta_{i,k,l}\right)\,,\end{eqnarray}
where 
$1\le i \le j-2$, $4\le j \le N$, $1\le k \le N-1$, and $1\le l \le N-1$  (therefore $i\leq N-2$). The triple delta function
$\delta_{i.j.k} = 1$ when $i=j=k$ and is zero otherwise.
Note that the above expressions for the Clebsch-Gordon coefficients
are valid for all $N$, not just a single $N$. In the context of this
project, this is a very important point. 

%

We could continue to derive the remaining coefficients in like manner, but due to the binary invariants and $W$ transformations involved in these sectors, the increase in complexity is prohibitive. It is most convenient at this point to use the above Clebsch-Gordon coefficients to derive those that remain. We derive $C^{221,\Rmn{1}}_{(ij)(kl)m}$ from
\begin{equation} C^{221,\Rmn{1}}_{(ij)(kl)m}= C^{211,\Rmn{1}}_{(ij)np}C^{211,\Rmn{1}}_{(kl)qr}C^{110,\Rmn{1}}_{pr}C^{111,\Rmn{1}}_{nqm}\end{equation}
where all of the implied summations run from $1$ to $N-1$ and where 
$1\le i \le j-2$, $4\le j \le N$, $1\le k \le l-2$, $4\le l \le N$ (therefore $i\leq N-2$ $k\leq N-2$).

The last two $C^{222,\mathcal{R}}$ Clebsch-Gordon coefficients $C^{222,\Rmn{1}}$, and $C^{222,\Rmn{2}}$ are a special case. Two linearly independent invariants may be formed as follows
\begin{equation} C^{222,\Rmn{1}}_{(ii')(jj')(kk')}= C^{211,\Rmn{1}}_{(ii')lm}C^{211,\Rmn{1}}_{(jj')np}C^{211,\Rmn{1}}_{(kk')qr}C^{110,\Rmn{1}}_{mn}C^{110,\Rmn{1}}_{pq}C^{110,\Rmn{1}}_{rl}\end{equation}
and
\begin{equation} C^{222,\Rmn{2}}_{(ii')(jj')(kk')}= C^{211,\Rmn{1}}_{(ii')lm}C^{211,\Rmn{1}}_{(jj')np}C^{211,\Rmn{1}}_{(kk')qr}C^{111,\Rmn{1}}_{lnq}C^{111,\Rmn{1}}_{mpr}\,,\end{equation}
where $1\le i \le j-2$, $4\le j \le N$, $1\le k \le l-2$, $4\le l \le N$,$1\le m \le n-2$, $4\le n \le N$. 
The Clebsch-Gordon coefficients $C^{221,\Rmn{1}}, C^{222,\Rmn{1}},
C^{222,\Rmn{2}}$ for arbitrary $N$ have a more complicated expression than the others which we do not give here. 
A \textsc{Mathematica}\cite{mathematica} package for all the required 
Clebsch-Gordon coefficients may be found in Ref.~\onlinecite{MATprog}.

\section{Multipliers $\beta({\mathcal G})$ of the Clebsch-Gordan Coefficients}\label{sec:bincoeff}
From Eq.~(\ref{eq:betaC}), the transformed binary invariants $B_W({\mathcal G})$ of Section~\ref{subsec:H12SC} are proportional to a Clebsch-Gordon coefficient of $S_N$ in the irreducible symmetry coordinate basis of Section~\ref{subsubsec:SCW} (a linear sum of two Clebsch-Gordon coefficients in the $\bm{222}$ sector). In this Appendix, we derive these proportionality coefficients $\beta({\mathcal G})$. These, together with the Clebsch-Gordon coefficients, will give us the symmetry coordinate transformed binary invariants, $B_{W}({\mathcal G})$\,, of Eqs.~(\ref{eq:E19}) and (\ref{eq:betaC}).

By design, the transformations $[W_X^{\alpha}]_{\nu,\eta}$ have a simpler expression for smaller values of the row index $\nu$\,. Also, the multipliers $\beta({\mathcal G})$ do not depend on the index $\nu$\,. Thus to derive $\beta^{\alpha_1
  \alpha_2 \alpha_3, \mathcal{R}} ({\mathcal G})$  from Eq.~(\ref{eq:betaC}), we use the lowest value for the index $\nu$ in Eqs.~(\ref{eq:W3beta}), (\ref{eq:W1beta}) (allowed by the row index restrictions on Eqs~(\ref{eq:WNm1r}) , (\ref{eq:WNm1g}), (\ref{eq:WNm2gijmn})), along with the Clebsch-Gordon coefficients of Appendix~\ref{sec:CCGC}, which yield a non-trivial equation (two coupled equations in the case of the $\bm{222}$ sector -- see below).
\subsection{Harmonic Order}
The only non-zero Clebsch-Gordon coefficients at harmonic-order which couple two orthogonal irreps.\ of $S_N$ together to form a scalar $[N]$ irrep.\ are proportional to $1$, $\mathbf{I}_{N-1}$, and $\mathbf{I}_{N(N-3)/2}$, and couple two $[N]$, $[N-1, \hspace{1ex} 1]$\,, and $[N-2, \hspace{1ex} 2]$ respectively  ($\mathbf{I}_n$ is the $n\times n$ unit matrix).  Thus we derive the proportionality coefficients from
\begin{equation}\label{eq:betatwoG}
[W^{\alpha}_{X_1}]_{\nu_1,\eta_1}[W^{\alpha}_{X_2}]_{\nu_2,\eta_2}[B^{X_1X_2}({\mathcal G})]_{\eta_1,\eta_2}=\betatwoG{\alpha}{\alpha}{\mathcal G}{X_1}{X_2}\delta_{\nu_1,\nu_2}\,.
\end{equation}
Choosing the lowest value of $\nu_1$, we obtain the results in Table~\ref{table:sigma2}.

\subsection{First Anharmonic Order}
The rank-one Clebsch-Gordon coefficient for a single $[N]$ irrep.\ to give an $[N]$ irrep.\ is obviously unity. Therefore, we obtain the proportionality coefficients from 
\begin{equation} \label{eq:betaoneG}
[W^{[N]}_{X}]_{\eta}[B({\mathcal G})]_{\eta}=\betaoneG{[N]}{\mathcal G}{X}\,,
\end{equation}
where $\mathcal G$ is $\graphr$ or $\graphgamma$\,. From Eq.~(\ref{eq:betaoneG}) we obtain the results in Table~\ref{table:sigma1}.]

The rank-three Clebsch-Gordon coefficients have a significantly more complicated form, but the proportionality coefficients $\betathreeG{\alpha_{1}}{\alpha_{2}}{\alpha_{3},\mathcal{R}}{\mathcal G}{X_1}{X_2}{X_3}$ may still be computed in closed form using a computer algebra system such as \textsc{Mathematica}\cite{mathematica} and our own packages. We solve Eq.~(\ref{eq:betaC}) for all but the $\bm{222}$ sector to give
\begin{equation*}\label{eq:betathreeG}
\betathreeG{\alpha_{1}}{\alpha_{2}}{\alpha_{3},\Rmn{1}}{\mathcal G}{X_1}{X_2}{X_3}=\frac{[W^{\alpha_{1}}_{X_1}]_{\nu_1,\eta_1}[W^{\alpha_{2}}_{X_2}]_{\nu_2,\eta_2}
[W^{\alpha_{3}}_{X_3}]_{\nu_3,\eta_3}[B({\mathcal G})]_{\eta_1,\eta_2,\eta_3}}{C^{\alpha_{1}\alpha_{2}\alpha_{3},\Rmn{1}}_{\nu_1,\nu_2,\nu_3}},
\end{equation*}
evaluated for the lowest $\nu_1,\nu_2,\nu_3$ (allowed by the restrictions on the indices of $W$) for which the Clebsch-Gordan coefficient is non-zero.

In the case of $\betathreeG{2}{2}{2,\Rmn{1}}{\mathcal G}{\gamma}{\gamma}{\gamma}$ and
$\betathreeG{2}{2}{2,\Rmn{2}}{\mathcal G}{\gamma}{\gamma}{\gamma}$, these may be evaluated from Eq.~(\ref{eq:betaC}),
i.e.\
\begin{eqnarray}\lefteqn{[B^{222}_{W}({\mathcal G})]_{(ij)(kl)(mn)}=}
\nonumber\\&&
\betathreeG{2}{2}{2 ,\Rmn{1}}{\mathcal G}{\gamma}{\gamma}{\gamma}[C^{222,\Rmn{1}}]_{(ij)(kl)(mn)}+\betathreeG{2}{2}{2 ,\Rmn{2}}{\mathcal G}{\gamma}{\gamma}{\gamma}[C^{222,\Rmn{2}}]_{(ij)(kl)(mn)}
\end{eqnarray}
for two different sets of ordered pair indices $(ij)(kl)(mn)$\,, where each of the ordered pairs are subject to the restrictions on the row indices of Eq~(\ref{eq:WNm2gijmn}).

We list all the multipliers, $\betathreeG{\alpha_{1}}{\alpha_{2}}{\alpha_{3},\mathcal{R}}{\mathcal G}{X_1}{X_2}{X_3}$\,,
of the Clebsch-Gordan coefficients in the expansion of the symmetry coordinate transformed, rank-three binary invariants,
$[B_W({\mathcal G})]_{\nu_1,\nu_2,\nu_3}$\,,  in Tables~\ref{tab:sigma3}-\ref{tab:tadpole}.

There is a subtlety regarding low values of $N$\,. When the number of vertices in a graph is less than $N$\,, the graph and corresponding binary invariant no longer exists (the vertices are associated with different values of the particle labels $i$\, $j$\, \ldots, etc.) Since the transformation from internal displacement coordinates to symmetry coordinates is a non-singular transformation, this reduction in the number of binary invariants is reflected in the number of non-zero, linearly independent Clebsch-Gordon coefficients at low $N$\,, i.e.\ the number is correspondingly reduced. These lower bounds on $N$ are noted in Tables~\ref{tab:sigma3}-\ref{tab:tadpole}.

In the case of the expansion of the derivative portion of the kinetic term there is yet another subtlety; it is necessary to distinguish between graph edges corresponding to derivatives and those corresponding to non-derivatives. In this case, we use a vertical ``tic'' to mark derivative edges. There is only one graph which requires this distinction: $\graphgggc$. The binary invariant $B(\graphgggc)$ is composed of the sum of binary invariants for the two graphs with two distinguishable edges: $B(\graphgggc)= B(\graphgggca)+ B(\graphgggcb)$. Only the binary invariant $B(\graphgggca)$ is present in the derivative term. The proportionality coefficient for this graph is shown in Table~\ref{tab:tadpole}.
%
%

\begin{thebibliography}{99}






\bibitem{liu:2007}
Y.~Liu, M.~Christandl, and F.~Verstraete,
\newblock Physical Review Letters {\bf 98}, 110503 (2007).

\bibitem{montina2008}
A.~Montina,
\newblock Physical Review A {\bf 77}, 022104 (2008).

\bibitem{dalfovo1999}
F. Dalfovo, S. Giorgini, L. Pitaevskii, and S. Stringari,
\newblock Rev. Mod. Phys. {\bf 71}, 463 (1999).


\bibitem{masiello2005}
See for example D.~Masiello, S.~McKagan, and W.~Reinhardt,
\newblock Physical Review A {\bf 72}, 63624 (2005).

\bibitem{CC} See for example L.S.\ Cederbaum, O.E.\ Alon, and A.I.\ Streltsov, Phys.\
Rev.\ A \textbf{73}, 043609 (2006); A.I.\ Streltsov, O.E.\ Alon, and L.S.\ Cederbaum, Phys. Rev. A \textbf{73}, 063626 (2006).

\bibitem{Reynolds:82} P.J.~Reynolds, D.M.~Ceperley, B.J.~Alder, and W.A.~Lester~Jr., J. Chem. Phys. \textbf{77}, 5593 (1982).

\bibitem{Hommond:94} B.L. Hammond, W.A. Lester, Jr., and P.J. Reynolds,
{\em Monte Carlo Methods in ab initio Quantum Chemistry}
(World Scientific, Singapore, 1994).

\bibitem{Bressanini:99} D. Bressanini and P.J. Reynolds, {\em Advances in Chemical Physics: Monte Carlo Methods in Chemical Physics} \textbf{105}, D.M. Ferguson, J.I. Siepmann, and D.G. Truhlar (Eds.), 37-64 (Wiley, New York, 1999).

\bibitem{dubois:01}
J.~L. DuBois and H.~R. Glyde,
\newblock Phys. Rev. A {\bf 63}, 023602 (2001).

\bibitem{dubois:03}
J.~L. DuBois and H.~R. Glyde,
\newblock Phys. Rev. A {\bf 68}, 033602 (2003).

\bibitem{purwanto:05}
W.~Purwanto and S.~Zhang,
\newblock Phys.\ Rev.\ A {\bf 72}, 053610 (2005).

\bibitem{mc} D.\ Landau and K.\ Binder,
{\it A Guide to Monte-Carlo Simulations in Statistical Physics},
(Cambridge University Press, Cambridge, 2001); M.\ Holzmann, W.\
Krauth, and M.\ Naraschewski, Phys.\ Rev.\ A \textbf{59}, 2956
(1999); J.K.\ Nilsen, J.\ Mur-Petit, M.\ Guilleumas, M.\
Hjorth-Jensen, and A.\ Polls, Phys.\ Rev.\ A \textbf{71}, 053610
(2005); D.\ Blume and C.H.\ Greene, Phys.\ Rev.\ A {\bf
63}, 063061 (2001).


\bibitem{MinguzziAnderson} A.\ Minguzzi, S.\ Succi, F.\ Toschi,
M.P.\ Tosi, and P.\ Vignolo, Phys.\ Rep.\ \textbf{395}, 223
(2004); J.O.\ Anderson, Rev.\ Mod.\ Phys.\ \textbf{76}, 599
(2004).




\bibitem{singh} A.\ Banerjee and M.\ P.\ Singh, Phys.\ Rev.\ A
  \textbf{64}, 063604 (2001).

\bibitem{nunes1999}
G.~Nunes,
\newblock Journal of Physics B: Atomic, Molecular, and Optical Physics 
{\bf 32}, 4293 (1999).


\bibitem{hamermesh} See for example M.\ Hamermesh, {\it Group Theory and its
Application to Physical Problems}, (Addison-Wesley, Reading, MA,
1962).

\bibitem{FGpaper} B.A.\ McKinney, M.\ Dunn, D.K.\ Watson, and J.G.\ Loeser,
  Ann.\ Phys.\ (NY) \textbf{310}, 56 (2003).

\bibitem{energy} B.A.\ McKinney, M.\ Dunn, D.K.\ Watson,
  Phys.\ Rev.\ A \textbf{69}, 053611 (2004).

\bibitem{paperI} W.B.\ Laing, M.\ Dunn, and D.K.\ Watson, Phys.\ Rev.\ A \textbf{74}, 063605 (2006).

\bibitem{zerothDensity} M.\ Dunn, D.K.\ Watson, and J.G.\ Loeser,
Ann.\ Phys.\ (NY), \textbf{321}, 1939 (2006).

\bibitem{Laing:arXiv:physics/0510177v1} W.~B. Laing, M.~Dunn, J.~G. Loeser,
and D.~K. Watson,
\newblock Arxiv preprint physics/0510177v1.

\bibitem{dcw} E.B.\ Wilson, Jr., J.C.\ Decius, P.C.\ Cross,
\textit{Molecular vibrations: The theory of infrared and raman
vibrational spectra}. McGraw- Hill, New York, 1955.

  \bibitem{quantumdots} See for example L.P.\ Kouwenhoven, D.G.\
Austing, and S.\ Tarucha, Rep.\ Prog.\ Phys.\ \textbf{64}, 701
(2001).

\bibitem{LegPit} See for example W.\ Ketterle, Rev.\ Mod.\ Phys.\ \textbf{74}, 1131 (2002);
E.A.\ Cornell and C.E.\ Wieman, Rev.\ Mod.\ Phys.\ \textbf{74},
875 (2002); A.J.\ Leggett, Rev.\ Mod.\ Phys.\ \textbf{73}, 307
(2001); L.\ Pitaevskii and S.\ Stringari, {\em Bose-Einstein
Condensation} (Oxford University Press, Oxford, 2003).

\bibitem{corbino} See for example P.\ Benetatos, and M.\ Manchetti, Phys.\ Rev.\ B
\textbf{65}, 134517 (2002).

\bibitem{rsfhs} See for example K.B.\ Whaley, in {it Advances
in Molecular Vibrations and Collision Dynamics}, vol.\ 3, J.\
Bowman, ed., JAI Press, Greenwich, Conn. (1998); J.P.\ Toennies,
A.F.\ Vilesov, and K.B.\ Whaley, Physics Today \textbf{54}, 31-37
(February 2001).





\bibitem{excitations} M.\ Dunn, W.B.\ Laing, and D.K.\ Watson, unpublished.


\bibitem{generaldpt} See for example 
J.G. Loeser, Z. Zhang, S. Kais, and D.R. Herschbach,
\newblock J. Chem Phys. {\bf 95}, 4525 (1991);
A. Holas, P.M. Kozlowski, and N.H. March,
\newblock J. Phys. A {\bf 24}, 4249 (1991);
C.M. Bender and S. Boettcher,
\newblock Phys. Rev. D {\bf 48}, 4919 (1993);
A. A. Svidzinsky, M.O. Scully, and D.R. Herschbach,
\newblock Phys. Rev. Lett. {\bf 95}, 080401 (2005);
A. Gonzalez,
\newblock Few-Body Systems {\bf 10}, 43 (1991);
A. Pagnamenta and U. Sukhatme,
\newblock Phys. Rev. D {\bf 34}, 3528 (1986).

\bibitem{copen92} {\em Dimensional Scaling in Chemical Physics},
  edited by D.R.\ Herschbach, J.\ Avery, and O.\ Goscinski (Kluwer,
  Dordrecht, 1992).

\bibitem{chattrev} A.\ Chatterjee, Phys.\ Rep.\ \textbf{186}, 249
  (1990).
  

\bibitem{wavefunction1harm}
W.~B. Laing, D.~W. Kelle, M.~Dunn, and D.~K. Watson,
\newblock Arxiv preprint math-ph/0902.3448v1; J.\ Phys.\ A, submitted.

\bibitem{density1harm}
M.~Dunn, W.~B. Laing,  D.~Toth, and D.~K. Watson,
\newblock Arxiv preprint cond-mat/0903.0875; Phys.\ Rev.\ A, submitted.

\bibitem{loeser} J.G.\ Loeser, J.\ Chem.\ Phys.\ \textbf{86}, 5635 (1987).

\bibitem{avery}J.\ Avery, D.Z.\ Goodson, D.R.\ Herschbach,
Theor.\ Chim.\ Acta \textbf{81}, 1 (1991).
\bibitem{covcont}Technically, the coordinates and derivatives
transform differently under $S_N$\,; contravariantly and covariantly
respectively (and denoted by superscripted and subscripted indices
respectively). Specifically, if the coordinate vector
${\bar{\bm{y}}}$ transforms contravariantly under a linear
transformation as ${\bar{\bm{y}}}_T = T \, {\bar{\bm{y}}}$ then the
derivative vector $\bm{\partial}_{\bar{\bm{y}}}$ transforms
covariantly as $\bm{\partial}_{\bar{\bm{y}}_T} =
 \bm{\partial}_{\bar{\bm{y}}} \, T^{-1}$\,.

Thus the $\bm{F}$ tensors are covariant, while the $\bm{G}$ tensors
are mixed tensors with two contravariant indices. However, the
internal displacement coordinates, ${\bar{\bm{y}}'}$ and symmetry
coordinates both transform under irreducible {\em orthogonal}
representations of the $S_N$ group. Thus in regards to the group
theoretic portion of the calculation, i.e.\ the derivation of the
symmetry coordinates, we drop the distinction between covariant and
contravariant tensors since, if $T$ is an element of the orthogonal
representation of $S_N$\,,
%
\begin{equation}
{[T^{-1}]^\rho}_\nu \, {[T]^\nu}_\varsigma = {[T^T]^\rho}_\nu \,
{[T]^\nu}_\varsigma = [g]^{\rho \xi} \, {[T]^\epsilon}_\xi \,
[g]_{\epsilon \nu} \,\, {[T]^\nu}_\varsigma = {[\delta]^\rho}_\nu
\,,
\end{equation}
%
where the covariant tensor $[g]_{\epsilon \nu}$ equals 1 when
$\epsilon=\nu$ and zero otherwise. Likewise for the contravariant
tensor $[g]^{\epsilon \nu}$\,. Two very important results follow
from this. First the metric tensor $[g]^{\epsilon \nu}$ is invariant
under $S_N$ since $[g]^{\epsilon \nu} = {[T]^\epsilon}_\varsigma \,
{[T]^\nu}_\xi \, [g]^{\varsigma \xi}$ \,. Likewise for
$[g]_{\epsilon \nu}$\,. Secondly
%
\begin{equation}
[\partial_{\bar{y}^\prime_T}]_\nu =
 [\partial_{\bar{y}^\prime}]_\rho \, {[T^{-1}]^\rho}_\nu =
 [\partial_{\bar{y}^\prime}]_\rho \, {[T^T]^\rho}_\nu =
 [\partial_{\bar{y}^\prime}]_\rho \, [g]^{\rho \xi} \, {[T]^\epsilon}_\xi \,
[g]_{\epsilon \nu} = [g]_{\nu \epsilon} \, {[T]^\epsilon}_\xi \,
[g]^{\xi \rho} \,\, [\partial_{\bar{y}^\prime}]_\rho \,,
\end{equation}
%
thus since $[g]_{\nu \epsilon} \, {[T]^\epsilon}_\xi \, [g]^{\xi
\rho}$ is numerically equal to ${[T]^\nu}_\rho$\,, under orthogonal
transformations covariant vectors transform in exactly the same fashion
as contravariant vectors. Thus in subsequent discussions regarding
the transformation to symmetry coordinates we will not distinguish
between covariant and contravariant indices. We will need though to
take into account the distinction between covariant and
contravariant vectors when we make the final transformation to
normal coordinates in Sec.~\ref{sec:transnorm}.
\bibitem{multigraph} Strictly speaking, this is a ``loop multigraph'': the definition of a graph does not allow for multiple
edges between a pair of vertices, nor a ``loop'' edge with common endpoints.
\bibitem{ongraphs}
See EPAPS Document No. for notation and a calculation of the binary invariants.
This document can be reached
  through a direct link in the online article’s HTML reference section or via
  the EPAPS homepage (http://www.aip.org/pubservs/epaps.html).
\bibitem{partition} See Appendix~B of Ref.~\onlinecite{higherL1}.
\bibitem{invariantC} See for example Ref.~\onlinecite{hamermesh}, Section~5-4, p.\ 136.
\bibitem{harmCG} The unit matrix $\bm{I_{\alpha}}$
is, up to a multiplier, the Clebsch-Gordon coefficient which couples two $\alpha$ irreps.\ together to form an $[N]$ irrep.\,.
\bibitem{MurnaghanGamba} F.D.\ Murnaghan, Am.\ J.\ Math., \textbf{60}, 761, (1938); A.\ Gamba and L. A.\ Radicati, Atti accad.\ nazl.\ Lincei, Rend.\ Classe sci.\ fis.\ mat.\ e nat., \textbf{14}, 632 (1953);
G.\ de Robinson and O.E.\ Taulbee,
Proc.\ Natl.\ Acad Sci.\ USA, \textbf{40}, 723 (1954); Ragey H.\ Makar, Proc.\ Edinburgh Math.\ Soc.\ \textbf{8}, 133-137 (1949).

\bibitem{fermcryst} See for example, D.\ Blume, Phys.\ Rev.\ A \textbf{66}, 053613 (2002);
I.\ Romanovsky, C.\ Yannouleas, U.\ Landman, Phys.\ Rev.\ Lett.\ \textbf{93}, 230405 (2004).

\bibitem{higherL1} M.\ Dunn and D.K.\ Watson, Ann.\
  Phys.\ (NY) \textbf{251}, 266 (1996).

\bibitem{higherL2} M.\ Dunn and D.K.\ Watson, Ann.\
  Phys.\ (NY) \textbf{251}, 319 (1996).

\bibitem{mathematica} I.\ Wolfram Research, \textsc{Mathematica} Edition: Version 6.0, 2007.





\bibitem{MATprog} W.B.\ Laing, M.\ Dunn, and D.K.\ Watson,
http://www.nhn.ou.edu/~watson/nbodydpt. The \textsc{Mathematica}
package for the Clebsch-Gordon coefficients is generated from the notebook
SNClebschGordon.nb



\end{thebibliography}
%

%
\newpage
\begin{table}[t]
\begin{tabular*}{0.7\textwidth}{@{\extracolsep{\fill}}cl}
\hline Graph & Matrix Elements\\\hline
 $\graphrra$ & $\stackleft{0}{2}Q^{rr}_{ii}$\\[0.7em]
 $\graphrrb$ & $\stackleft{0}{2}Q^{rr}_{ij}$\\
 $\graphgra$ & $\stackleft{0}{2}Q^{r\gamma}_{i(ij)}(=\stackleft{0}{2}Q^{r\gamma}_{i(ji)}),
  \stackleft{0}{2}Q^{\gamma r}_{(ij)i}(=
 \stackleft{0}{2}Q^{\gamma r}_{(ij)j})$\\
 $\graphgrb$ & $\stackleft{0}{2}Q^{r\gamma}_{i(jk)} , \stackleft{0}{2}Q^{\gamma r}_{(ij)k}$\\
 $\graphgga$ & $\stackleft{0}{2}Q^{\gamma \gamma}_{(ij)(ij)}$\\
$\graphggb$ & $\stackleft{0}{2}Q^{\gamma \gamma}_{(ij)(ik)}(=\stackleft{0}{2}Q^{\gamma \gamma}_{(ij)(jk)}), \stackleft{0}{2}Q^{\gamma \gamma}_{(ij)(ki)}(=\stackleft{0}{2}Q^{\gamma \gamma}_{(ij)(kj)})$\\
$\graphggc$ & $\stackleft{0}{2}Q^{\gamma \gamma}_{(ij)(kl)}$\\
\hline
\end{tabular*}
\caption{Graphs for rank-two
tensors at harmonic order, along with the corresponding tensor elements.}
\label{tab:ranktwo}
\end{table}
\begin{table}[t]
\begin{tabular*}{0.4\textwidth}{@{\extracolsep{\fill}}cc}
\hline Graph & Tensor Elements\\\hline
 $\graphr$ & $\stackleft{1}{1}Q^{r}_{i}$\\
 $\graphgamma$ & $\stackleft{1}{1}Q^{\gamma}_{(ij)}$\\
\hline
\end{tabular*}
\caption{Graphs for rank-one
tensors at first-order beyond harmonic, along with the corresponding tensor elements.}
\label{tab:rankone}
\end{table}
%
%
\begin{table}[p]
\begin{tabular}{|cl|cl|cl|}
\hline
Graph & Tensor Elements &Graph & Tensor Elements &Graph & Tensor Elements\\
\hline
 $\graphrrra$ & $\stackleft{1}{3}Q^{rrr}_{i,i,i}$
	&$\graphggra$ & $\stackleft{1}{3}Q^{\gamma \gamma r}_{(ij),(ij),i}$
	&$\graphggga$ & $\stackleft{1}{3}Q^{\gamma \gamma \gamma}_{(ij),(ij),(ij)}$\\
 $\graphrrrb$ & $\stackleft{1}{3}Q^{rrr}_{i,i,j}$
	&$\graphggrb$ & $\stackleft{1}{3}Q^{\gamma \gamma r}_{(ij),(ij),k}$
	&$\graphgggb$ & $\stackleft{1}{3}Q^{\gamma \gamma \gamma}_{(ij),(jk),(ik)}$\\
 $\graphrrrc$ & $\stackleft{1}{3}Q^{rrr}_{i,j,k}$
	&$\graphggrc$ & $\stackleft{1}{3}Q^{\gamma \gamma r}_{(ij),(jk),j}$
	&$\graphgggc$ & $\stackleft{1}{3}Q^{\gamma \gamma \gamma}_{(ij),(ij),(jk)}$\\
 \cline{1-2}
 $\graphgrra$ & $\stackleft{1}{3}Q^{\gamma rr}_{(ij),i,i}$
	&$\graphggrd$ & $\stackleft{1}{3}Q^{\gamma \gamma r}_{(ij),(jk),i}$
	&$\graphgggd$ & $\stackleft{1}{3}Q^{\gamma \gamma \gamma}_{(ij),(jk),(jl)}$\\
 $\graphgrrb$ & $\stackleft{1}{3}Q^{\gamma rr}_{(ij),i,j}$
	&$\graphggre$ & $\stackleft{1}{3}Q^{\gamma \gamma r}_{(ij),(jk),l}$
	&$\graphggge$ & $\stackleft{1}{3}Q^{\gamma \gamma \gamma}_{(ij),(jk),(kl)}$\\
 $\graphgrrc$ & $\stackleft{1}{3}Q^{\gamma rr}_{(ij),i,k}$
	&$\graphggrf$ & $\stackleft{1}{3}Q^{\gamma \gamma r}_{(ij),(kl),i}$
	&$\graphgggf$ & $\stackleft{1}{3}Q^{\gamma \gamma \gamma}_{(ij),(ij),(kl)}$\\
 $\graphgrrd$ & $\stackleft{1}{3}Q^{\gamma rr}_{(ij),k,k}$
	&$\graphggrg$ & $\stackleft{1}{3}Q^{\gamma \gamma r}_{(ij),(kl),m}$
	&$\graphgggg$ & $\stackleft{1}{3}Q^{\gamma \gamma \gamma}_{(ij),(jk),(lm)}$\\
 $\graphgrre$ & $\stackleft{1}{3}Q^{\gamma rr}_{(ij),k,l}$
	&&
	&$\graphgggh$ & $\stackleft{1}{3}Q^{\gamma \gamma \gamma}_{(ij),(kl),(mn)}$\\
\hline
\end{tabular}
\caption{Graphs for rank-three tensors at first-order beyond harmonic, along with the corresponding tensor elements.}
\label{tab:rankthree}
\end{table}
\begin{table}
\centering
\begin{displaymath}
 \begin{array}[c]{|c|cc|}
\hline
 & \mathbf{00} & \mathbf{11}  \\
& 1\leq N & 2\leq N
\\\hline
 \graphrra & 1 & 1  \\
 \graphrrb & N-1 & -1  \\
\hline
\end{array}
\end{displaymath}
\begin{displaymath}
 \begin{array}[c]{|c|cc|}
\hline
 & \mathbf{00} & \mathbf{11} \\
& 2\leq N & 3\leq N
\\\hline
  \graphgra & \sqrt{2(N-1)} & \sqrt{N-2}   \\
 \graphgrb & \frac{1}{2}(N-2)\sqrt{2(N-1)}  & -\sqrt{N-2}   
\\
\hline
\end{array}
\end{displaymath}
\begin{displaymath}
 \begin{array}[c]{|c|ccc|}
\hline
 & \mathbf{00} & \mathbf{11} & \mathbf{22} \\
& 2\leq N & 3\leq N & 4 \leq N
\\\hline
 \graphgga & 1 & 1 & 1 \\
 \graphggb & 2 (N-2) & N-4 & -2 \\
 \graphggc & \frac{1}{2} (N-3) (N-2) & -(N-3) & 1
 \\\hline
\end{array}
\end{displaymath}
\caption{Proportionality coefficients of the harmonic-order transformed binary invariants} 
\label{table:sigma2}
\end{table}
%
\begin{table}
\centering
\begin{tabular}[c]{|c|c@{\hspace{4ex}}c@{\hspace{2ex}}@{\hspace{2ex}}c@{\hspace{2ex}}|}
\hline
 & $\mathbf{0}$ & $\mathbf{1}$ & $\mathbf{2}$ \\
 \hline
 \graphr & $\sqrt{N}$ & 0 & 0 \\
 \graphgamma & $\sqrt{\frac{N(N-1)}{2}}$ & 0 & 0
\\ \hline
\end{tabular}
\caption{Proportionality coefficients of the first anharmonic order transformed binary invariants} 
\label{table:sigma1}
\end{table}
\begin{table}[htbp]
\begin{displaymath}
\begin{array}[c]{|c|ccc|}
\hline&&&\\
\betathreeG{\alpha_1}{\alpha_2}{\alpha_3,\mathcal{R}}{\mathcal G}{r}{r}{r}& \mathbf{000} & \mathbf{110},\mathbf{101},\mathbf{011} & \mathbf{111} 
\\
&(1\leq N)&(2\leq N)&(3 \leq N)
\\
   \hline
 \graphrrra&\frac{1}{\sqrt{N}}& \frac{1}{\sqrt{N}} & 1 \\
 \graphrrrb& \frac{3 (N-1)}{\sqrt{N}} & \frac{N-3}{\sqrt{N}} & -3 \\
 \graphrrrc& \frac{(N-2) (N-1)}{\sqrt{N}} & \frac{2-N}{\sqrt{N}} & 2 
\\
\hline
\end{array}
\end{displaymath}
\caption{Multipliers of the Clebsch-Gordon coefficients of the first-anharmonic-order transformed binary invariants: $\betathreeG{\alpha_1}{\alpha_2}{\alpha_3\mathcal,{R}}{\mathcal G}{r}{r}{r} $}
\label{tab:sigma3}
\end{table}
\begin{table}[ht]
\begin{displaymath}
\begin{array}[c]{|c|ccccc|}
\hline&&&&&\\
\betathreeG{\alpha_1}{\alpha_2}{\alpha_3,\mathcal{R}}{\mathcal G}{\gamma}{r}{r}& \mathbf{000} & \mathbf{110},\mathbf{101}& \mathbf{011} & \mathbf{111} &\mathbf{211} 
\\
&(2\leq N)&(2\leq N)&(3\leq N)&(3\leq N)&(4\leq N)
\\\hline
 \graphgrra& 2\sqrt{\frac{N-1}{2N}} & \sqrt{\frac{N-2}{N}} & \frac{\sqrt{2} \sqrt{N - 1}}{\sqrt{N}} & \sqrt{N-2}  &0\\
 \graphgrrb& 2\sqrt{\frac{N-1}{2N}} & \sqrt{\frac{N-2}{N}} & -\frac{\sqrt{2}}{\sqrt{N - 1} \sqrt{N}} & -\frac{2}{\sqrt{N-2}} &1\\
 \graphgrrc& 4\sqrt{\frac{N-1}{2N}} & (N-4) \sqrt{\frac{N-2}{N}} & -\frac{2 \sqrt{2} (N - 2)}{\sqrt{N - 1} \sqrt{N}} & -\frac{2 (N-4)}{\sqrt{N-2}} &-2\\
 \graphgrrd& (N-2)\sqrt{\frac{N-1}{2N}} & -\sqrt{\frac{N-2}{N}} & \frac{(N - 2) \sqrt{N - 1}}{\sqrt{2} \sqrt{N}} & -\sqrt{N-2} &0\\
 \graphgrre&\frac{1}{2}{}_NP_4\sqrt{\frac{2}{N^3(N-1)}}
 & -(N-3) \sqrt{\frac{N-2}{N}} & \frac{(3 - N) (N - 2)}{\sqrt{2} \sqrt{N - 1} \sqrt{N}} & \frac{2 (N-3)}{\sqrt{N-2}} &1
\\\hline
\end{array}
\end{displaymath}
\caption{Multipliers of the Clebsch-Gordon coefficients of the first-anharmonic-order transformed binary invariants: $\betathreeG{\alpha_1}{\alpha_2}{\alpha_3\mathcal,{R}}{\mathcal G}{\gamma}{r}{r} $\,, where ${}_NP_4 = N!/(N-4)!$}
\label{table:sigma3grr}
\end{table}
\begin{turnpage}
\begin{table}[htbp]
\begin{tabular}{c}
$ \begin{array}[c]{|c|ccccccc|}
\hline&&&&&&&\\
 \betathreeG{\alpha_1}{\alpha_2}{\alpha_3,\mathcal{R}}{G}{\gamma}{\gamma}{r}& \mathbf{000} & \mathbf{110}&\mathbf{011},\mathbf{101}& \mathbf{111} & \mathbf{211},\mathbf{121} &\mathbf{220}&\mathbf{221}
\\
&(2\leq N)&(3\leq N)&(3\leq N)&(3\leq N)&(4\leq N)&(4\leq N)&(5\leq N)
\\
\hline
 {\graphggra}& \frac{2}{\sqrt{N}} & \frac{2}{\sqrt{N}} &\sqrt{2} \sqrt{\frac{N - 2}{(N - 1) N}}& \frac{N-4}{N-2} & \frac{1}{\sqrt{N-2}} & \frac{2}{\sqrt{N}} & \frac{1}{\sqrt{N-2}} \\
 {\graphggrb}& \frac{N-2}{\sqrt{N}} & \frac{N-2}{\sqrt{N}} &-\sqrt{2} \sqrt{\frac{N - 2}{(N - 1) N}}& \frac{4-N}{N-2} & -\frac{1}{\sqrt{N-2}} & \frac{N-2}{\sqrt{N}} & -\frac{1}{\sqrt{N-2}} \\
 {\graphggrc}& \frac{2 (N-2)}{\sqrt{N}} & \frac{N-4}{\sqrt{N}} &\sqrt{2} (N - 2) \sqrt{\frac{N - 2}{(N - 1) N}}& \frac{N^2-5 N+8}{N-2} & -\frac{1}{\sqrt{N-2}} & -\frac{2}{\sqrt{N}} & -\frac{1}{\sqrt{N-2}}
   \\
 {\graphggrd}& \frac{4 (N-2)}{\sqrt{N}} & \frac{2 (N-4)}{\sqrt{N}} &\sqrt{2} (N - 4) \sqrt{\frac{N - 2}{(N - 1) N}}& \frac{16-6 N}{N-2} & \frac{N-4}{\sqrt{N-2}} & -\frac{4}{\sqrt{N}} &
   -\frac{2}{\sqrt{N-2}} \\
 {\graphggre}& \frac{2 (N-3) (N-2)}{\sqrt{N}} & \frac{(N-4) (N-3)}{\sqrt{N}} &-2 \sqrt{2} (N - 3) \sqrt{\frac{N - 2}{(N - 1) N}}& -\frac{(N-8) (N-3)}{N-2} & \frac{5-N}{\sqrt{N-2}} & -\frac{2 (N-3)}{\sqrt{N}}
   & \frac{3}{\sqrt{N-2}} \\
 {\graphggrf}& \frac{2 (N-3) (N-2)}{\sqrt{N}} & -\frac{4 (N-3)}{\sqrt{N}} &(N - 4) (N - 3) \sqrt{\frac{N - 2}{2(N - 1) N}}& -\frac{2 (N-4) (N-3)}{N-2} & \frac{4-N}{\sqrt{N-2}} & \frac{4}{\sqrt{N}} &
   \frac{2}{\sqrt{N-2}} \\
 {\graphggrg}& \frac{(N-4) (N-3) (N-2)}{2 \sqrt{N}} & -\frac{(N-4) (N-3)}{\sqrt{N}} &-(N - 4) (N - 3) \sqrt{\frac{N - 2}{2(N - 1) N}}& \frac{2 (N-4) (N-3)}{N-2} & \frac{N-4}{\sqrt{N-2}} &
   \frac{N-4}{\sqrt{N}} & -\frac{2}{\sqrt{N-2}} 
\\\hline
\end{array} $
\end{tabular}
\caption{Proportionality coefficients of the first-anharmonic-order transformed binary invariants: $\betathreeG{\alpha_1}{\alpha_2}{\alpha_3\mathcal,{R}}{\mathcal G}{\gamma}{\gamma}{r} $}
\label{table:sigma3ggr}
\end{table}
\end{turnpage}
\begin{table}[htbp]
\begin{scriptsize}
\begin{displaymath}
\begin{array}[c]{|c|cccc|}
\hline&&&&\\
  \betathreeG{\alpha_1}{\alpha_2}{\alpha_3,\mathcal{R}}{\mathcal G}{\gamma}{\gamma}{\gamma} & \mathbf{000} & \mathbf{110},\mathbf{101},\mathbf{011}& \mathbf{111} & \mathbf{211},\mathbf{121},\mathbf{112}
\\
&(2\leq N)&(3\leq N)&(3\leq N)&(4\leq N)
\\
   \hline
 {\graphggga}& \frac{1}{2}{}_NP_2 \left(\frac{2}{N(N-1)}\right)^\frac{3}{2} & \frac{\sqrt{2}}{\sqrt{(N-1) N}} & \frac{N-4}{(N-2)^{3/2}} &\frac{1}{N-2}
\\
 {\graphgggb}& {}_NP_3 \left(\frac{2}{N(N-1)}\right)^\frac{3}{2} & \frac{\sqrt{2} (N-4)}{\sqrt{(N-1) N}} & \frac{16-6 N}{(N-2)^{3/2}} & \frac{N-4}{N-2} 
\\
 {\graphgggc}& 3{}_NP_3 \left(\frac{2}{N(N-1)}\right)^\frac{3}{2} & \frac{4 \sqrt{2} (N-3)}{\sqrt{(N-1) N}} & \frac{3 (N-4)^2}{(N-2)^{3/2}} & \frac{2 (N-5)}{N-2} 
\\
 {\graphgggd}& {}_NP_4 \left(\frac{2}{N(N-1)}\right)^\frac{3}{2} & \frac{\sqrt{2} (N-4) (N-3)}{\sqrt{(N-1) N}} & \frac{(N-3) \left(N^2-6
   N+16\right)}{(N-2)^{3/2}} & -\frac{2 (N-4)}{N-2} \\
 {\graphggge}& 3{}_NP_4 \left(\frac{2}{N(N-1)}\right)^\frac{3}{2} & \frac{2 \sqrt{2} (N-6) (N-3)}{\sqrt{(N-1) N}} & -\frac{12 (N-4) (N-3)}{(N-2)^{3/2}} &
   \frac{(N-6) (N-5)}{N-2} \\
 {\graphgggf}& \frac{3}{4}{}_NP_4 \left(\frac{2}{N(N-1)}\right)^\frac{3}{2} & \frac{(N-6) (N-3)}{\sqrt{2} \sqrt{(N-1) N}} & -\frac{3 (N-4) (N-3)}{(N-2)^{3/2}} &
   \frac{7-2 N}{N-2} \\
 {\graphgggg}&\frac{3}{2}{}_NP_5 \left(\frac{2}{N(N-1)}\right)^\frac{3}{2} & \frac{(N-12) (N-4) (N-3)}{\sqrt{2} \sqrt{(N-1) N}} & -\frac{3 (N-8) (N-4)
   (N-3)}{(N-2)^{3/2}} & -\frac{(N-4) (2 N-13)}{N-2} \\
 {\graphgggh}& \frac{1}{8}{}_NP_6 \left(\frac{2}{N(N-1)}\right)^\frac{3}{2} & -\frac{(N-5) (N-4) (N-3)}{\sqrt{2} \sqrt{(N-1) N}} & \frac{2 (N-5) (N-4)
   (N-3)}{(N-2)^{3/2}} & \frac{(N-5) (N-4)}{N-2} \\\hline
\end{array}
\end{displaymath}
\begin{displaymath}
\begin{array}[c]{|c|ccccc|}
\hline&&&&&\\
\betathreeG{\alpha_1}{\alpha_2}{\alpha_3,\mathcal{R}}{\mathcal G}{\gamma}{\gamma}{\gamma}   & \mathbf{220},\mathbf{202},\mathbf{022} &
  \mathbf{221},\mathbf{212},\mathbf{122} & \mathbf{222}&\mathbf{222(i)} & \mathbf{222(ii)} 
\\
&(4\leq N)&(5\leq N)&(4\leq N\leq 5)&(6\leq N)&(6\leq N)
\\
   \hline
 {\graphggga}& \sqrt{\frac{2}{N(N-1)}} & \frac{1}{\sqrt{N-2}} &\frac{1}{4}& 0 & \frac{1}{4} \\
 {\graphgggb}& -2
   \sqrt{\frac{2}{N(N-1)}} & -\frac{2}{\sqrt{N-2}} &1& 1 & 0 \\
 {\graphgggc}&
   2 (N-4) \sqrt{\frac{2}{N(N-1)}} & \frac{N-8}{\sqrt{N-2}} &-\frac{3}{2}& 0 & -\frac{3}{2} \\
 {\graphgggd}& -2  (N-3) \sqrt{\frac{2}{N(N-1)}} & \frac{6-N}{\sqrt{N-2}} &1& 0 & 1 \\
 {\graphggge}& -4 (N-4)\sqrt{\frac{2}{N(N-1)}} & -\frac{2 (N-8)}{\sqrt{N-2}} &-\frac{3}{2}& -3 & \frac{3}{2} \\
 {\graphgggf}& \frac{1}{2}\left(N^2-5 N+10\right) \sqrt{\frac{2}{N(N-1)}} & \frac{5-N}{\sqrt{N-2}} &\frac{3}{4}& 0 & \frac{3}{4} \\
 {\graphgggg}& -(N-7) (N-4) \sqrt{\frac{2}{N(N-1)}} & \frac{5 N-28}{\sqrt{N-2}} &0& 3 & -3 \\
 {\graphgggh}& \frac{1}{2}(N-5) (N-4) \sqrt{\frac{2}{N(N-1)}} & -\frac{2 (N-5)}{\sqrt{N-2}} &0& -1 & 1 \\ \hline
\end{array}
\end{displaymath}
\end{scriptsize}
\caption{Proportionality coefficients of the first-anharmonic-order transformed binary invariants: $\betathreeG{\alpha_1}{\alpha_2}{\alpha_3,\mathcal{R}}{\mathcal G}{\gamma}{\gamma}{\gamma} $\,, where ${}_NP_i = N!/(N-i)!$}
\label{tab:sigma3ggg}
\end{table}
\begin{table}[thbp]
\begin{scriptsize}
\begin{displaymath}
\begin{array}[c]{|c|cccccc|}
\hline &&&&&&\\
\betathreeG{\alpha_1}{\alpha_2}{\alpha_3,\mathcal{R}}{\mathcal G}{\gamma}{\gamma}{\gamma}   & \mathbf{000} & \mathbf{110},\mathbf{101}&\mathbf{011}& \mathbf{111} & \mathbf{211}&\mathbf{121},\mathbf{112}
\\
&(2\leq N)&(3\leq N)&(3\leq N)&(3\leq N)&(4\leq N)&(4\leq N)
\\
   \hline
\graphgggca& 2{}_NP_3 \left(\frac{2}{N(N-1)}\right)^\frac{3}{2} 
&\frac{\sqrt{2} (3
   N-8)}{\sqrt{(N-1) N}}
&\frac{2 \sqrt{2}
   (N-4)}{\sqrt{(N-1) N}}
& \frac{3
   (N-4)^2}{(N-2)^{3/2}}
&2-\frac{4}{N-2}
&\frac{N-6}{N-2}
\\
\hline
\end{array}
\end{displaymath}
\begin{displaymath}
\begin{array}[c]{|c|ccccccc|}
\hline &&&&&&&\\
\betathreeG{\alpha_1}{\alpha_2}{\alpha_3,\mathcal{R}}{\mathcal G}{\gamma}{\gamma}{\gamma} & \mathbf{022}&\mathbf{202},\mathbf{220} 
  &\mathbf{122}&\mathbf{221},\mathbf{212} & \mathbf{222}&\mathbf{222(i)} & \mathbf{222(ii)} 
\\
&(4\leq N)&(4\leq N)&(5\leq N)&(5\leq N)&(4\leq N\leq 5)&(6\leq N)&(6\leq N)
\\
   \hline
 \graphgggca
&-\frac{4 \sqrt{2}}{\sqrt{(N-1)
   N}}
&\frac{2 \sqrt{2}
   (N-3)}{\sqrt{(N-1) N}}
&-\frac{4}{\sqrt{N-2}}
&\frac{N-6}{\sqrt{N-2}}
&-1
&0
&-1
\\\hline
\end{array}
\end{displaymath}
\end{scriptsize}
\caption{Proportionality coefficients of the first-anharmonic-order transformed binary invariants with distinguishable edges: relevant graphs}
\label{tab:tadpole}
\end{table}
\end{document}